\documentclass[journal]{IEEEtran}
\usepackage[usenames,dvipsnames]{color}
\usepackage{setspace}
\usepackage{amsmath}
\usepackage{graphicx}
\usepackage{epstopdf}
\usepackage{amssymb}
\usepackage[hyphens]{url}
\usepackage{cite}
\usepackage{subfig}
\usepackage[usenames,dvipsnames]{color}
\usepackage{setspace}
\usepackage{bm}
\usepackage[utf8]{inputenc}
\usepackage[english]{babel}
\usepackage{amsthm}
\usepackage{enumerate}
\usepackage{siunitx}
\usepackage[export]{adjustbox}

\usepackage{mathdots}
\usepackage{balance}
\usepackage{algorithm}
\usepackage{listings}
\usepackage[keeplastbox]{flushend}
\usepackage{fancybox}
\usepackage{epsfig}
\usepackage{mathtools}

\newcommand{\comm}{\mathrm{c}}

\newcommand{\Es} {{\mathcal{E}_{\mathrm{s}}} }

\newcommand{\e}[1]{{\mathbb E}\left[ #1 \right]}
\newcommand{\me}{\mathrm{e}}
\newcommand{\eff}{\mathrm{eff}}

\newcommand{\jm}{\mathrm{j}}

\newcommand{\mr}{\mathrm{r}}

\newcommand{\mtar}{k}

\newcommand{\mrad}{\mathrm{r}}
\newcommand{\mcom}{\mathrm{c}}

\newcommand{\Ts}{T_{\mathrm{s}}}
\newcommand{\Tc}{T_{\mathrm{s}}}

\newcommand{\txm}{\mathrm{t}}

\newcommand{\thm}{\mathrm{th}}

\newcommand{\um}{0}

\def\bb0{{\mathbb{0}}}

\def\bb{{\mathbf{b}}}

\def\bee{{\mathbf{e}}}
\def\bff{{\mathbf{f}}}

\def\bh{{\mathbf{h}}}

\def\bw{{\mathbf{w}}}

\def\by{{\mathbf{y}}}
\def\bz{{\mathbf{z}}}
\def\b0{{\mathbf{0}}}

\def\bA{{\mathbf{A}}}

\def\bH{{\mathbf{H}}}

\def\bJ{{\mathbf{J}}}

\def\bU{{\mathbf{U}}}

\def\bZ{{\mathbf{Z}}}

\def\sf0{{\mathsf{0}}}

\def\rmD{\mathrm{D}}

\def\rmS{\mathrm{S}}
\def\rmT{\mathrm{T}}

\def\rmc{{\mathrm{c}}}

\def\rmp{{\mathrm{p}}}

\def\rm0{{\mathrm{0}}}

\usepackage{mathtools}
\usepackage[]{hyperref}
\hypersetup{pdftex,colorlinks=true,allcolors=blue}
\usepackage{hypcap}

\newcommand{\imj}{\mathsf{j}}

\newcommand{\bzc}{\boldsymbol{\xi}}
\newcommand{\fftn}{q}
\newcommand{\quant}{b}
\newcommand{\optr}{\rho}
\newcommand{\dftf}{\tilde{\bff}}
\allowdisplaybreaks

\newcommand{\dmseeff}{\mathrm{DMSE}_\eff}
\newcommand{\nmse}{\mathrm{NMSE}}
\newcommand{\Tr}[1]{\mathrm{Tr}\left[ #1 \right]}
\newcommand{\cvx}[1]{\mathrm{Conv}\left( #1 \right)}

\usepackage{amsmath,graphicx}

\usepackage{eqparbox}
\usepackage{algorithmic}

\usepackage{lipsum}
\usepackage{fancyhdr}
\usepackage{datetime}

\begin{document}

\title{ Adaptive and Fast Combined Waveform-Beamforming Design for mmWave Automotive Joint Communication-Radar}

\author{{Preeti~Kumari, Nitin~Jonathan~Myers, and~Robert~W.~Heath,~Jr.}
\thanks{ Preeti Kumari, Nitin Jonathan Myers, and Robert W. Heath Jr. are with the Wireless Networking and Communications Group, the University of Texas at Austin, TX 78712-1687, USA (e-mail: \{preeti\_kumari, nitinjmyers, rheath\}@utexas.edu).

This material is based upon work supported in part by the National Science Foundation under Grant No. ECCS-1711702 and in part by the Army Research Office under grant W911NF1910221.}}

{}

\maketitle

\begin{abstract}
Millimeter-wave (mmWave) joint communication-radar (JCR) will enable high data rate communication and high-resolution radar sensing for applications such as autonomous driving. Prior JCR systems that are based on the mmWave communications hardware, however, suffer from a limited angular field-of-view and low estimation accuracy for radars due to the employed directional communication beam. In this paper, we propose an adaptive and fast combined waveform-beamforming design for the mmWave automotive JCR with a phased-array architecture that permits a trade-off between communication and radar performances. To rapidly estimate the mmWave automotive radar channel in the Doppler-angle domain with a wide field-of-view, our JCR design employs a few circulant shifts of the transmit beamformer and apply two-dimensional partial Fourier compressed sensing technique. We optimize these circulant shifts to achieve minimum coherence in compressed sensing. We evaluate the JCR performance trade-offs using a normalized mean square error (MSE) metric for radar estimation and a distortion MSE metric for data communication, which is analogous to the distortion metric in the rate-distortion theory. Additionally, we develop a MSE-based weighted average optimization problem for the adaptive JCR combined waveform-beamforming design. Numerical results demonstrate that our proposed JCR design enables the estimation of short- and medium-range radar channels in the Doppler-angle domain with a low normalized MSE, at the expense of a small degradation in the communication distortion MSE.
\end{abstract}


\IEEEpeerreviewmaketitle

\section{Introduction}
Millimeter-wave (mmWave) communication and radar are key technologies for many next-generation applications, such as autonomous driving. MmWave automotive radars provide high-resolution sensing with a wide field of view (FoV)~\cite{PatTorWan:Automotive-Radars:-A-review:17}, while mmWave communications will enable a high data rate solution for the next-generation connected vehicles~\cite{ChoVaGon:Millimeter-Wave-Vehicular-Communication:16}.  The combination of these two technologies into a single joint communication-radar (JCR) enables hardware reuse and a common signaling waveform. This leads to benefits in power consumption, spectrum efficiency, and market penetrability. Unfortunately, a fully-digital multiple-input-multiple-output mmWave JCR with high-speed, high-resolution analog-to-digital converters will result in huge power consumption and high hardware complexity due to the wide available bandwidth and high dimensions.

To mitigate these issues, we propose an adaptive and fast combined waveform-beamforming design for the mmWave automotive JCR that uses a phased-array architecture. Such an architecture is used in the mmWave WLAN IEEE 802.11ad standard~\cite{ieee2012wireless}. In this design, we exploit all the transmit (TX) antennas during the data transmission mode to generate a narrow coherent beam for communication and constant gain sidelobes in other directions for radar sensing. Then, the transmitter applies a few circulant shifts of the designed TX beamformer, and employs two-dimensional (2D) partial Fourier compressed sensing (CS) technique to rapidly estimate the Doppler-angle domain radar channel.  To achieve better radar channel reconstruction with CS, we optimize the sequence of circulant shifts used at the TX. Additionally, we use a generic JCR TX waveform structure with tunable preamble length to increase the range for automotive radar sensing. Then, we develop a mean square error (MSE)-based adaptive combined waveform-beamforming design for the mmWave automotive JCR to find the optimal mainlobe gain for communication and the optimal preamble length to achieve high channel estimation accuracy for medium-range radar (MRR) and short-range radar (SRR), at the cost of a small reduction in the vehicle-to-vehicle (V2V) communication rate.

Most prior work on mmWave automotive JCR systems are either radar-centric or communication-centric~\cite{MisBhaKoi:Toward-Millimeter-Wave-Joint:19,JRC2020}. In the radar-centric JCR, the communication messages are modulated on top of the radar waveforms~\cite{DokShaSti:Multicarrier-Phase-Modulated:18}, or the communication information is embedded in the TX beamforming vectors~\cite{HasAmiZha:Signaling-strategies-for-dual-function:16}. These systems, however, do not support high data rates as the communication signal must be spread to avoid disturbing the radar required properties and they employ analog pre-processing in the time-domain. In \cite{KumChoGon:IEEE-802.11ad-Based-Radar::18}, a communication-centric mmWave automotive JCR with fully-digital time-domain processing was developed by exploiting the preamble of the IEEE 802.11ad standard~\cite{ieee2012wireless}. Using simulations, it was shown that IEEE 802.11ad-based JCR can simultaneously achieve high range/velocity resolution for automotive long-range radar (LRR) sensing and gigabits-per-second data rates for V2V communications. The IEEE 802.11ad standard, however, supports single-stream analog beamforming that leads to a large trade-off between communication and radar performances.

Prior approaches to increase the radar FoV for mmWave automotive JCR can be categorized into three types: (a) JCR during the communication beam training mode, (b) JCR with an adaptive beamforming design during the data transmission mode, and (c) multiple-input-multiple-output JCR with low resolution analog-to-digital converters. In the first approach~\cite{GroLopVen:Opportunistic-Radar-in-IEEE:18,MunMisGue:Beam-Alignment-and-Tracking:19}, the communication beam training mode was proposed for radar sensing. In~\cite{GroLopVen:Opportunistic-Radar-in-IEEE:18}, the IEEE 802.11ad beam scanning algorithm was exploited for radar detection/estimation with a wide FoV. In~\cite{MunMisGue:Beam-Alignment-and-Tracking:19}, a new MAC configuration for vehicle-to-infrastructure JCR application was proposed that employed beam switching pattern with dedicated sectors for radar and communication. In the second approach~\cite{KumEltHea:Sparsity-aware-adaptive-beamforming:18}, the IEEE 802.11ad SC PHY frames along with the adaptive random switching (RS) of TX antennas during the data transmission mode was proposed. In the RS-JCR, a coherent beam is formed towards the communication receiver, while simultaneously perturbing the grating lobes of the resulting beam pattern for angle-of-arrival (AoA) estimation in SRR applications.  In the last approach~\cite{KumMazMez:Low-Resolution-Sampling-for-Joint:18}, a mmWave multiple-input-multiple-output JCR with 1-bit analog-to-digital converters per RF chain was proposed to achieve a high range and AoA estimation accuracy. The RS-JCR has a higher radar update rate than the first approach, and is based on a commercially available mmWave hardware unlike the third approach. The RS-JCR, however, employs TX antenna subsets instead of using all antennas, which decreases the net TX power for JCR operation under a per-antenna power constraint. Additionally, the RS-JCR was developed for SRR channel estimation in the angular domain only.

In this paper, we develop an adaptive combined waveform-beamforming design for mmWave automotive JCR that exploits all the TX antennas during the data transmission mode to perform a highly accurate SRR/MRR Doppler-angle domain channel estimation, at the cost of a small reduction in the communication data rate. We assume that the location and relative velocity of a target remain constant during a coherent processing interval (CPI). This is justified by the small enough acceleration and velocity of a target relative to the radar sensor, as found in automotive applications~\cite{PatTorWan:Automotive-Radars:-A-review:17}. We also assume full-duplex radar operation due to the recent development of systems with sufficient isolation and self-interference cancellation~\cite{LiJosTao:Feasibility-study-on-full-duplex:14}. Lastly, we assume perfect data interference cancellation on the training part of the received JCR waveform because the transmitted data is known at the radar receiver, similar to~\cite{KumVorHea:Adaptive-Virtual-Waveform:20}. The main contributions of this paper are summarized as follows:
\begin{itemize}
\item  We propose a novel formulation for a mmWave automotive JCR system that performs automotive MRR and SRR sensing in a wide FoV without reducing the communication data rate much. This formulation captures the nuances of the sparse mmWave JCR channel with multiple targets in the Doppler-angle domain. Our proposed JCR system employs a generic TX waveform structure and uses a tunable TX beamforming design that can be optimized to achieve enhanced JCR performance using sparse sensing techniques. 

\item We develop a convolutional CS (CCS)-JCR technique to estimate the 2D-radar channel in the Doppler-angle domain. In this technique, the transmitter applies fewer circulant shifts of the JCR TX beamformer to acquire distinct CS measurements at the radar receiver. We transform our CS problem into a partial Fourier CS problem in 2D~\cite{Rau:Compressive-sensing-and-structured:10}. We show that the space-time sensing constraints in our problem allows only fewer configurations of the subsampling locations in partial Fourier CS.

\item We propose an optimized CCS (OCCS)-JCR approach by carefully designing the circulant shifts applied at the transmitter for superior Doppler-angle domain channel reconstruction. The optimized circulant shifts result in a space-time sampling pattern that achieves minimum coherence in partial Fourier CS under the space-time sensing constraints of our developed JCR system model.

\item We investigate the JCR performance trade-off using the NMSE metric for radar and a comparable DMSE metric for communication using analysis and simulations. Additionally, we formulate a MSE-based weighted average optimization problem for an adaptive combined waveform-beamforming mmWave JCR design in automotive applications that meets the Pareto-optimal bound. We solve the MSE-based optimization problem for our proposed OCCS-JCR approach in different target scenarios.

\item Numerical results demonstrate that the proposed OCCS-JCR combined waveform and beamforming design estimates the MRR and SRR radar channel in the Doppler-angle domain with low NMSE, at the cost of small reduction in the communication DMSE. The proposed OCCS-JCR performs the best, followed by the random CCS (RCCS)-JCR that uses random circulant shifts, and the RS-JCR extended for the Doppler-angle domain radar channel estimation performs the worst.

\end{itemize}

The work in this paper is a significant extension of our submitted conference papers~\cite{KumEltHea:Sparsity-aware-adaptive-beamforming:18,KumMyeVor:A-Combined-Waveform-Beamforming-Design:19}. In addition to the detailed exposition, we have included Doppler effect in the CCS-JCR system model, joint Doppler-angle estimation, optimized space-time sampling pattern, adaptive combined waveform and beamforming design, a MSE-based weighted average optimized JCR design, and numerical results to demonstrate and evaluate the performance of our proposed CCS-JCR design.

The rest of this paper is organized as follows. We formulate a JCR system model with a generic TX waveform structure and beamformer design algorithm in Section~\ref{sec:SystemModel}. In Section~\ref{sec:CCS}, we describe the CCS algorithm to estimate the radar channel in the Doppler-angle domain. Then, we outline the space-time sampling pattern optimization for the CCS-JCR in Section~\ref{sec:subOpt}. In Section~\ref{sec:AdapJCR}, we describe the performance metrics and adaptive combined-waveform beamforming design for the proposed CCS-JCR. We present the numerical results in Section~\ref{sec:Results}. Finally, we conclude our work and provide direction for future work in Section~\ref{sec:Conclusion}.

\textbf{Notation}: The operators $(\cdot)^*$ stands for conjugate transpose and $(\cdot)^{\mathrm{T}}$ for transpose of a matrix or a vector. $\mathcal{N}(\mu,\sigma^2)$ is used for a complex circularly symmetric Gaussian random variable with mean $\mu$ and variance $\sigma^2$. The set of integers is represented by $\mathbb{Z}$ and the set of real numbers is represented by $\mathbb{R}$. For a vector $\mathbf{a}$, $\mathbf{a}^k$ is a vector in which every entry of $\mathbf{a}$ is raised to the power of $k$. $\mathbf{A} \odot \mathbf{B}$ is defined as the element-wise multiplication of $\mathbf{A}$ and $\mathbf{B}$. We use $\bee_{m,M} \in \mathbb{C}^{M \times 1}$ to represent the $m^\thm$ standard basis vector for Euclidean space of real numbers. We use $\mathrm{phase}_\quant(x)$ to denote the $\quant$-bit phase quantized version of $x$. The matrix $\mathbf{U}_N \in \mathbb{C}^{N \times N}$ denotes the unitary Discrete Fourier Transform (DFT) matrix.

\section{System model} \label{sec:SystemModel}
%
\begin{figure}[!t]
\centering
\includegraphics[width=\columnwidth]{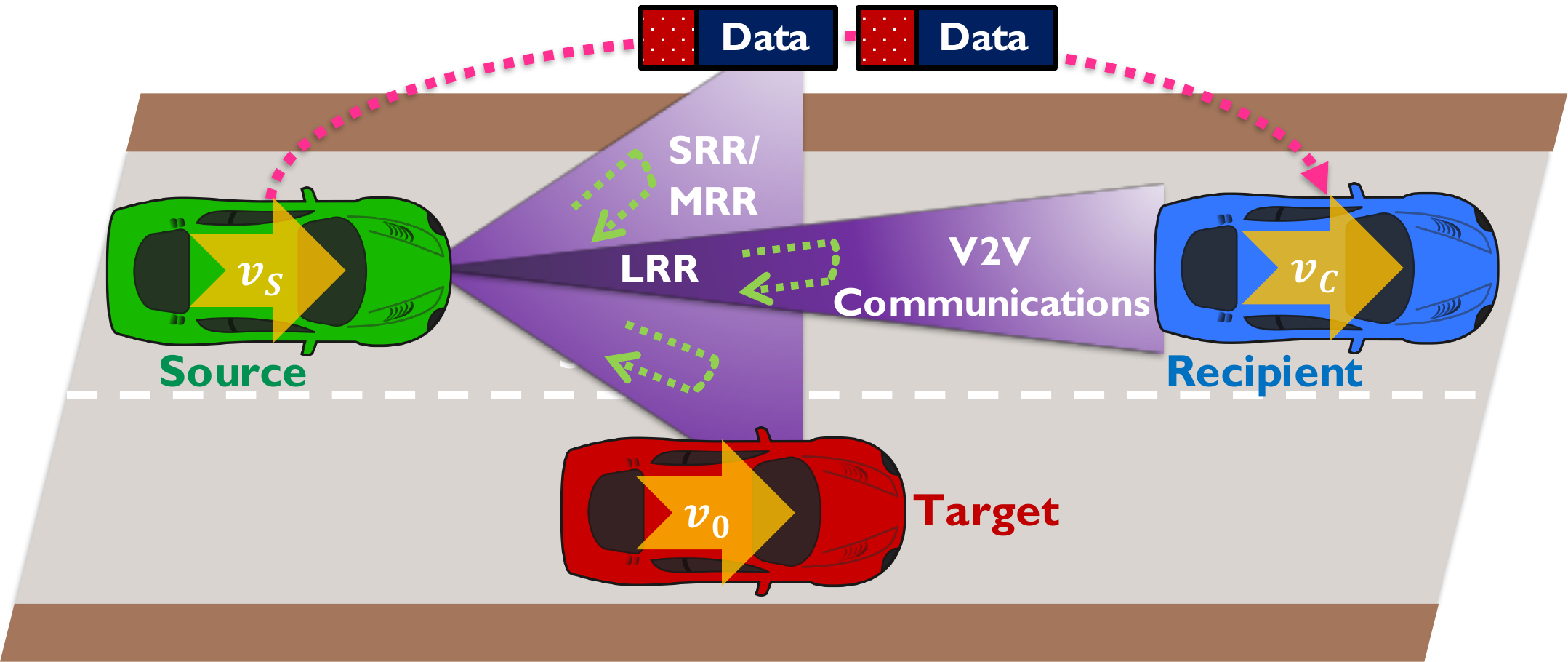}
 \caption{An illustration of an automotive mmWave JCR system that simultaneously perform SRR/MRR radar sensing with a wide FoV and V2V communication with a narrow FoV.}
\label{fig:scenario}
\vspace{-1.2em}
\end{figure}

We consider the use case where a source vehicle sends a mmWave JCR waveform to communicate with a recipient vehicle at a distance $d_\comm$ moving with a relative velocity $v_\comm$, while simultaneously using the received echoes for automotive radar sensing, as shown in Fig.~\ref{fig:scenario}. We consider closely separated TX antenna array and RX antenna array mounted on both source and recipient vehicles. For simplicity, we assume that the antenna arrays are uniform linear arrays (ULAs) with $N$-elements each. We assume a phased array architecture with $\quant$-bit phase shifters at the TX and the RX, where the phase shift alphabet is defined as $\mathbb{Q}_\quant=\{e^{\imj 2 n \pi /2^\quant}/ \sqrt{N}: n\in \{1,2,3,\cdots 2^b\} \}$. We assume the JCR transmitter generates a narrow beam towards the communication receiver without any blockage and distributes the remaining energy uniformly across the other directions for radar sensing.

\subsection{Waveform design for JCR} \label{sec:beamdesign}

We consider a CPI of $T$ seconds. We consider a generic TX waveform structure with $M$ equi-spaced frames separated by an inter-frame space (IFS) of $T_\mathrm{IFS}$. Each $L$-symbols frame consists of a $L_\mrad$-symbols preamble part and $(L-L_\mrad)$-symbols communication data segment. We assume that the training sequences possess good correlation properties for communication channel estimation, and the training sequence length is an integer multiple $\rho$ of the building block size $L_\mathrm{BLK} = L_\mrad/\rho$. The mmWave WLAN standard~\cite{ieee2012wireless} with Golay complementary sequences can realize this JCR preamble structure. Additionally, the IEEE 802.11ad standard can realize this multi-frame approach using the block/no acknowledgment policy during the communication between a dedicated pair of nodes in the data transmission interval~\cite[Ch. 9]{ieee2012wireless}. Similar to~\cite{KumChoGon:IEEE-802.11ad-Based-Radar::18}, we exploit the training sequences used in the preamble with good properties for radar sensing. 

To unambiguously estimate a maximum relative target velocity $v_\mathrm{max}$ in a CPI, the $m^\thm$ frame is considered to be located at an integer multiple, $m$, of the Doppler Nyquist sampling interval $T_\rmD$. Here, $T_\rmD \leq \lambda /(4 v_\mathrm{max})$. To enhance the radar estimation performance of the mmWave JCR without decreasing communication rate much, we propose to optimize $\optr$ and thereby the length of the training sequence. The analysis in this paper can be extended to a virtual waveform design structure with non-uniformly placed frames similar to~\cite{KumVorHea:Adaptive-Virtual-Waveform:20}.

\begin{figure}[!t]
\centering
\includegraphics[width=\columnwidth]{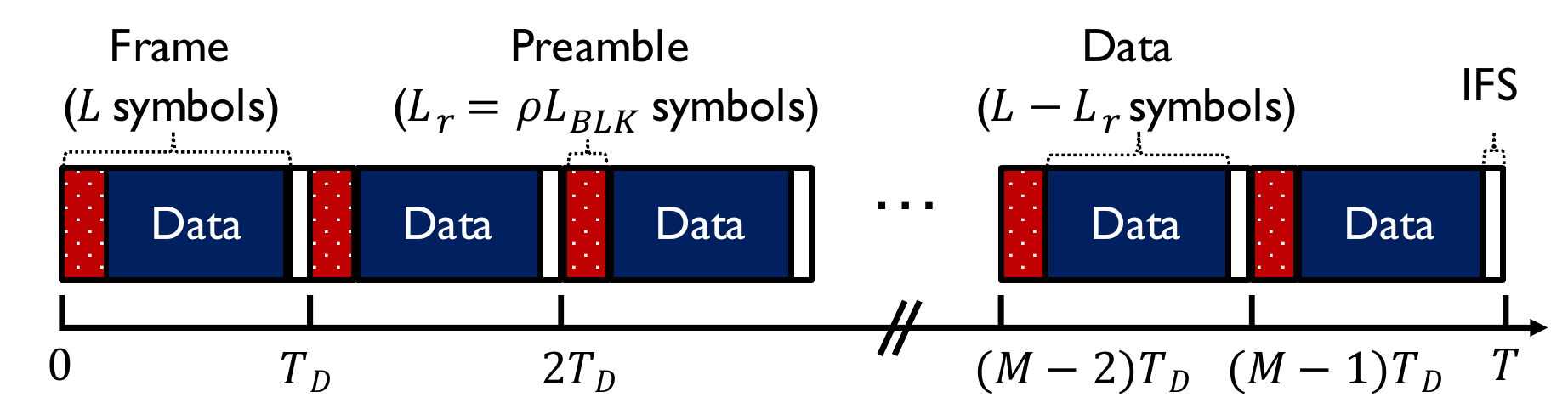}
 \caption{A CPI of $T$ seconds duration with $M$ JCR equi-spaced frames separated by an IFS of $T_\mathrm{IFS}$. Each $L$-symbols frame consists of $L_\mrad = \rho L_\mathrm{BLK}$ number of preamble symbols and $(L-L_\mrad)$ number of communication data symbols. Each frame is placed at an integer multiple of $T_\rmD$.}
\label{fig:Frame_Uni}
\vspace{-1.4em}
\end{figure}

We denote the unit energy TX pulse-shaping filter as $g_\txm(t)$, the signaling bandwidth as $W$, and the symbol period as $\Tc \approx 1/W$. The $\ell^\thm$ TX symbol corresponding to the $m^\thm$ frame is denoted by $s_{m,\ell}[\optr]$, which satisfies the average power constraint $\e{\vert s_{m,\ell}[\optr] \vert^2 } = \Es$. Then, the generic complex-baseband continuous-time representation of the single-carrier TX waveform in a CPI is given as
\begin{equation} \label{eq:TXCont}
x(t,\optr) = \sum_{m =0}^{M -1}  \sum_{\ell=0}^{L-1} s_{m,\ell}[\optr] g_\txm(t -\ell\Tc-mT_\rmD),
\end{equation}
where $L = (T_\rmD-T_\mathrm{IFS})/\Ts$.

\subsection{Beamformer design for JCR} \label{sec:beamdesign}

\par In this subsection, we explain our approach to construct a collection of beamformers that are well suited to the automotive JCR application. Our method first constructs one sequence for each of the TX and RX. These sequences are designed according to the JCR specification. In this specification, a good JCR beamformer is one that has a reasonable gain along the communication direction and sufficient power along the sensing directions for radar. The use of a single JCR beamformer, however, may not be sufficient to detect multiple targets. To this end, our method constructs a collection of beamformers by circularly shifting the beamformers constructed at the TX and the RX. As circulant shifts of a vector preserve the magnitude of its DFT, the proposed method ensures that the beams constructed according to our procedure achieve the desired JCR specification. The collection of circularly shifted beamformers is used to acquire distinct radar channel measurements.

\begin{figure}[!ht]
\begin{minipage}[b]{\linewidth}
  \centering
  \centerline{ \includegraphics[clip,scale=0.65]{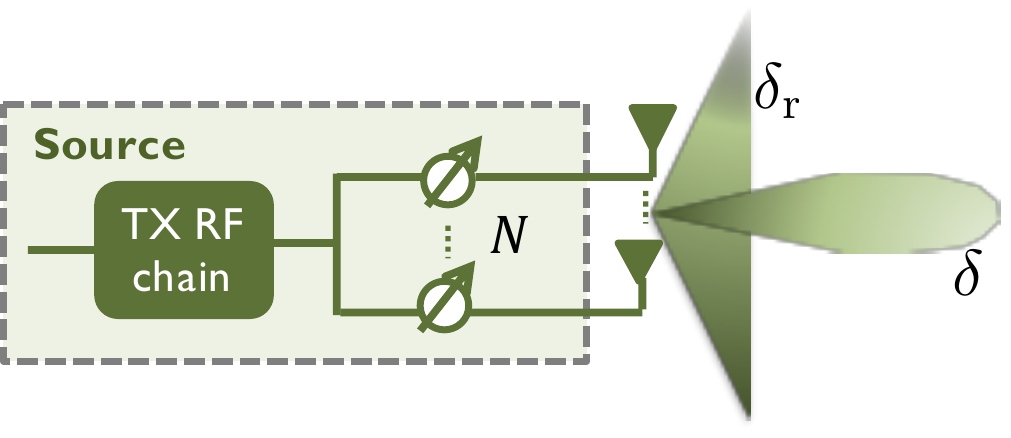}}
  \centerline{(a) JCR TX beamforming at the source vehicle}\medskip
\end{minipage}
\begin{minipage}[b]{\linewidth}
  \centering
  \centerline{\includegraphics[clip,scale=0.65]{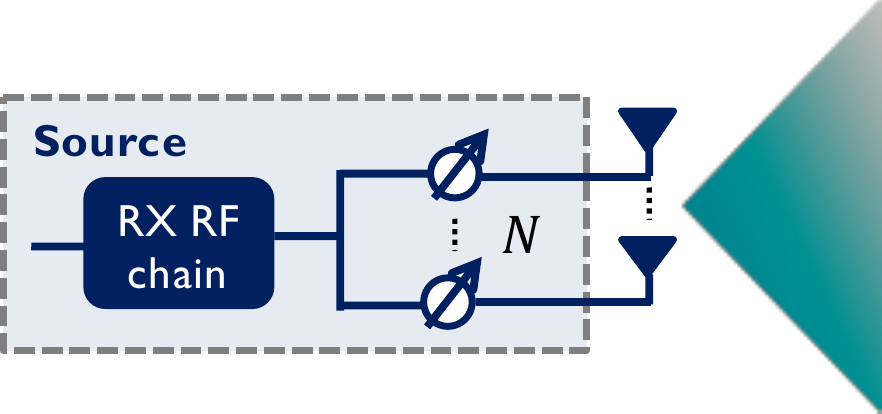}}
  \centerline{(b) Radar RX beamforming at the source vehicle}\medskip
\end{minipage}
\begin{minipage}[b]{\linewidth}
  \centering
   \vspace{0.2cm}
  \centerline{\includegraphics[clip,scale=0.65]{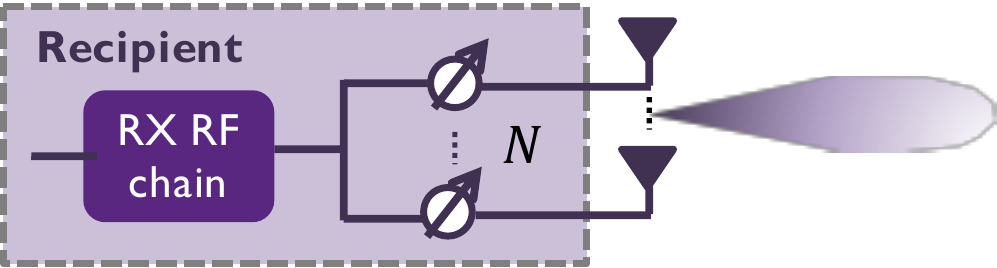}}
 \vspace{0.2cm}
  \centerline{(c) Communication RX beamforming at the recipient vehicle}\medskip
\end{minipage}
\caption{The JCR TX ULA at the source vehicle in (a) uses all the antennas to generate a narrow coherent beam for communication and distribute the remaining energy uniformly along the other directions for radar sensing.  The radar RX ULA at the source vehicle in (b) forms a constant gain beam for radar sensing, while the communication RX ULA at the recipient vehicle in (c) generates a narrow coherent beam pointed towards the JCR source transmitter. }
\label{fig:TXRXbeam}
\vspace{-1.2em}
\end{figure}

During a CPI, our JCR design uses an adaptive collection of TX and RX beams to achieve a high-resolution radar sensing in a wide FoV with a minimal reduction in the communication data rate. We propose to use $\delta$ fraction of TX power along the communication receiver direction, $\theta_0$, and $1-\delta$ fraction of TX power along the other directions for radar sensing. Within a CPI, the transmitter applies $M$ different beamforming vectors to acquire distinct radar channel measurements. The source vehicle uses $M$ unit norm TX beamforming vectors $\{ \mathbf{f}_m(\delta) \}_{m=0}^{M-1}$. 
Therefore, the TX signal at the source vehicle during a CPI is
\begin{equation} \label{eq:tx_beam}
{\mathbf{x}_{\txm}}(t,\optr) = \bff_m(\delta) \text{ } x(t,\optr), \quad 0 \leq t \leq T .
\end{equation}
To enhance the radar estimation performance of the mmWave JCR without decreasing communication rate much, we also propose to optimize $\delta$ under a per-antenna power constraint.

\par Now, we explain the key idea underlying the proposed TX beamformer design technique. For tractability, we design the beamformer by considering a DFT grid with $N$ discrete angles. For ease of exposition, we assume the communication direction is $0^\circ$. The JCR TX beamformer design problem is to design a sequence $\bff_\txm(\delta) \in \mathbb{Q}^N_\quant$ whose beampattern has an energy of $\delta$ along $0^\circ$. The remaining energy in the beamformer must be distributed across the other directions to enable radar channel reconstruction with fewer channel measurements. Prior work has shown that beamformers with close to uniform gain along the desired sensing directions enable fast CS channel reconstruction~\cite{MyeMezHea:FALP:-Fast-beam:19}. To this end, the proposed construction distributes the energy of $1-\delta$ ``uniformly'' across the remaining DFT grid locations, as shown in Fig.~\ref{fig:TXRXbeam}.

\par  We use the Gerchberg Saxton (GS) algorithm~\cite{Ger:A-practical-algorithm-for-the-determination:72} to construct the desired JCR beamformer at the TX [See Algorithm 1]. By the JCR specification, we seek an $\bff_\txm(\delta)$ whose discrete beam pattern has an energy of $\delta$ along $0^{\circ}$ and $\delta_r=(1-\delta)/(N-1)$ along the remaining $N-1$ directions. The discrete beam pattern is simply the $N$-point DFT of the vector $\bff_\txm(\delta)$. Therefore, the DFT magnitude vector associated with the desired beamformer is 
\begin{equation} \label{eq:DFTmagf}
\dftf_{\mathrm{mag}}(\delta) = [\sqrt{\delta},\sqrt{\delta_\mrad},\sqrt{\delta_\mrad},\cdots,\sqrt{\delta_\mrad}]^\rmT. 
\end{equation}
A naive approach to construct the desired JCR beamformer is to apply an inverse DFT over $\dftf_{\mathrm{mag}}(\delta)$. The inverse DFT of $\dftf_{\mathrm{mag}}(\delta)$, however, may not be an element in $\mathbb{Q}^{N}_\quant$. The GS algorithm is an alternating projection method that finds a sequence in $\mathbb{Q}^{N}_\quant$ such that the magnitude of its DFT is close to $\dftf_{\mathrm{mag}}(\delta)$.

\begin{algorithm} \label{algo:GS}
\caption{GS algorithm to find $\bff_\txm(\delta)$}
\label{GAMP_alg}
\begin{algorithmic}[1]
\STATE \textbf{Inputs:}  $\delta$, $N$, $\quant$, $T_{\mathrm{GS}}$, and $\dftf_{\mathrm{mag}}(\delta)$.
\STATE \textbf{Initialize:} Set $t_{\mathrm{iter}}=1$ and $\bff_\txm(\delta)$ to a Zadoff-Chu sequence.   \\    
\WHILE{$t_{\mathrm{iter}}<T_{\mathrm{GS}}$}
\STATE $\dftf_{\mathrm{phase}} (\delta)\leftarrow \mathrm{phase}\left( \mathrm{DFT}(\bff_\txm(\delta)) \right)$ 
\STATE Constraint on the discrete beam pattern:\\${\dftf(\delta)} \leftarrow
        \dftf_{\mathrm{mag}}(\delta) \odot \mathrm{exp}(\imj \dftf_{\mathrm{phase}}(\delta))$
\STATE Constraint on the antenna weights:\\ $\mathbf{f}_{\mathrm{phase}}(\delta) \leftarrow \mathrm{phase}_\quant\left( \mathrm{IDFT}(\dftf(\delta))\right)$
\STATE $\bff_\txm(\delta) \leftarrow \mathrm{exp}(\imj \mathbf{f}_{\mathrm{phase}}(\delta))/\sqrt{N}$
\ENDWHILE 
\RETURN $\bff_\txm(\delta)$.
\end{algorithmic}
\end{algorithm}

The proposed GS-based beamformer design procedure can be generalized for any communication direction $\theta \neq 0^{o}$. For an $N$-element ULA with elements half-wavelength spaced, we define the array steering vector $\textbf{a}(\theta) \in \mathbb{C}^{N \times 1}$ as 
\begin{equation}
\textbf{a}(\theta) = \left[1,e^{\imj \pi \mathrm{sin} \theta},e^{\imj 2 \pi \mathrm{sin} \theta},\cdots,e^{\imj (N-1) \pi \mathrm{sin} \theta}\right]^\rmT.
\end{equation} 
The TX beamformer in such a case is defined as $\bff_\txm(\delta) \odot \mathbf{a}(\theta)$. The JCR beamformer $\mathbf{f}_m(\delta)$ is constructed by circulantly shifting $\bff_\txm(\delta)$. As there are $N$ distinct circulant shifts of the $N$-length vector $\bff_\txm(\delta)$, there are $N$ candidates for $\mathbf{f}_m(\delta)$ with our design. The TX beam pattern achieved by the GS algorithm for $N = 256$ and its comparison with RS-based JCR as well as ideal communications are illustrated in our paper~\cite{KumMyeVor:A-Combined-Waveform-Beamforming-Design:19}.

The receiver at the source vehicle uses a unit norm RX beamforming vector $\bff_\mrad$, and the receiver at the recipient vehicle employs a unit norm beamforming vector $\bff_{\mcom}$. For the radar receiver at the source vehicle, a good CS-based beamformer is one that has equal energy at all DFT-grid locations within a desired sector~\cite{MyeMezHea:FALP:-Fast-beam:19}. We propose to use a ZC sequence in $\mathbb{Q}^N_q$ to be the RX  combiner vector $\bff_\mrad$ because the DFT of a ZC sequence has a constant amplitude. For the communication receiver at the recipient vehicle, we use a spatial matched filter of the communication channel as the RX beamformer $\bff_{\mcom}$ to provide the maximum TX-RX array gain and thereby achieve the highest communication spectral efficiency.
\subsection{Received signal model}
Within a CPI of $T$ seconds, we assume that the acceleration and the relative velocity of a moving target is small enough to assume constant velocity and that the target is quasi-stationary (constant location parameters). After the RX matched filtering, and symbol rate sampling, the communication/radar RX signal model in a CPI can be formulated as follows.

\emph{Communication received signal model:} To explore the performance trade-off between communication and radar, we consider an illustrative example of a line-of-sight, frequency-flat mmWave communication channel between the source and recipient vehicles~\cite{KumEltHea:Sparsity-aware-adaptive-beamforming:18}. Nonetheless, the approach developed in this paper can be extended for different scattering scenarios by including frequency-selective communication channels; the extension is omitted because of space limitations. We assume that the channel is time-invariant during a single frame because the relative velocity between the source and target vehicles are small. We do not include band-limited filters in the channel model and instead include them in the TX/RX signal models. The communication channel between the source and recipient vehicle is characterized by its complex channel amplitude $h_\mcom$, angle-of-departure (AoD) and AoA pair $(\theta_\um,\phi_\um)$, path delay $d_\comm/c$ with $c$ being the speed of light. 

Assuming perfect synchronization and additive noise ${w}_{\rmc,m,\ell} \sim  \mathcal {N}(0,\sigma_\mcom^2)$, the received communication signal with the TX steering vector $\mathbf{a}(\theta_\um)$, the RX steering vector $\mathbf{a}(\phi_\um)$, and the channel $\bH_\mcom = {h_\mcom}\mathbf{a}(\phi_\um)\mathbf{a}^*(\theta_\um)$ is
\begin{eqnarray}\label{eq:commRx1}
{y}_{\rmc,m,\ell}(\optr,\delta) =  \bff_{\mcom}^* \bH_\mcom  \mathbf{f}_m(\delta)s_{m,\ell}[\optr] + {w}_{\rmc,m,\ell}.
\end{eqnarray}
Assuming that the TX and RX beams are perfectly aligned and directional beamforming with a spatial matched filter is used at the RX to provide the maximum TX-RX array gain for the considered line-of-sight channel model, (\ref{eq:commRx1}) simplifies as
\begin{eqnarray}\label{eq:commRx2}
{y}_{\rmc,m,\ell}(\optr,\delta) =  \sqrt{ \delta }N {h}_\mcom s_{m,\ell}[\optr] + {w}_{\rmc,m,\ell}.
\end{eqnarray}
We define communication signal-to-noise ratio (SNR) corresponding to the ideal beampattern for communication with $\delta = 1$ as ${\zeta}_\mcom=\Es \vert {h}_\mcom \vert^2 N^2/{\sigma_\rmc}^2$. In this case, the net received signal SNR increases linearly with the fraction of TX power for communication and is given by $ \delta {\zeta_\mcom}$.


\emph{Radar received signal model:} We represent the doubly selective (time- and frequency-selective) mmWave radar channel using virtual representation obtained by uniform sampling in range, Doppler, and AoD dimensions \cite{KumChoGon:IEEE-802.11ad-Based-Radar::18}. 
Since the focus of this paper is target detection/estimation in the Doppler-AoD domain and not in the range domain, we describe radar signal model for a particular dominant range bin with distance $d$~\cite{KumEltHea:Sparsity-aware-adaptive-beamforming:18}.  The same algorithm can be applied to each range bin.

We assume that the range bin of interest consists of a few, $K$, virtual target scattering centers. The $\mtar^{\mathrm{th}}$ virtual scattering center is described by its Doppler-AoD pair $(\nu_k,\theta_{\mtar})$ and complex channel amplitude $h_{\mtar}$, which is a product of radar cross-section and path-loss. After the RX beamfoming, after the cross-correlation of the TX training sequences with the $m^\thm$ received frame echo, and assuming perfect cancellation of the data part on the received training signal~\cite{KumVorHea:Adaptive-Virtual-Waveform:20}, the radar received signal corresponding to the training part with an additive noise ${w}_{m}[\rho]$ is given as
\begin{eqnarray}\label{eq:rady}
y_m(\optr,\delta) =   \sum_{k=0}^{K-1} {h_k} \me^{- \jm 2\pi \nu_{k} m T_\rmD} \mathbf{a}^*(\theta_k) \mathbf{f}_m(\delta) +  w_m[\optr].
\end{eqnarray}
We assume the thermal noise in the receiver is an additive white Gaussian noise with variance $\sigma^2$. We denote $\gamma[\optr]$ as the product of the RX beamforming gain and the integration gain due to the employed cross-correlation, which depends on the training sequence length used within a frame. Then, the additive noise $w_m[\optr]$ in \eqref{eq:rady} is distributed as $\mathcal {N}(0, \sigma^2/{\Es  \gamma[\optr] })$.

We denote the Doppler shift vector $ \mathbf{d}(\nu_k)$ as
\begin{equation}\label{eq:Dopvec}
 \mathbf{d}(\nu_k) = \left[ 1, \me^{- \jm 2\pi \nu_{k} T_\rmD}, \cdots, \me^{- \jm 2\pi \nu_{k} (M-1)T_\rmD} \right] ^\rmT
\end{equation}
and the $m^\thm$ standard basis vector of length $M$ as $\bee_{m,M} \in \mathbb{C}^{M\times 1}$ with $\bee^\rmT_{m,M} = [ 0, \dots, 1, \cdots, 0]$, where $1$ is at the $(m+1)^\thm$ place. For example, $\bee^\rmT_{0,M}=[1,0,0,\cdots,0]$. We represent the radar channel in a CPI be expressed as 
\begin{equation}\label{eq:H}
\mathbf{H} = \sum_{k=0}^{K-1} {h_k} \mathbf{d}(v_k) \mathbf{a}^*(\theta_k).
\end{equation}
We observe that  $\me^{- \jm 2\pi \nu_{k} m T_\rmD}$ in \eqref{eq:rady} can be expressed as $\bee^\rmT_{m,M}\mathbf{d}(\nu_k)$. Putting this observation in \eqref{eq:rady} and using the definition of $\mathbf{H}$, we can write
\begin{eqnarray}\label{eq:yradFinal}
y_m(\optr,\delta) =  \bee^\rmT_{m,M} \mathbf{H}\mathbf{f}_m(\delta) +  w_m[\optr].
\end{eqnarray}
We define the SNR of the received radar signal excluding the preamble correlation gain and the TX beamforming gain as $\zeta = {\Es \beta \gamma[0]}/\sigma^2$ with average target channel power $\beta$. We denote the SNR that includes the preamble correlation gain but excludes the TX beamforming gain as $\zeta_\rmp[\rho] = {\Es \beta \gamma[\optr]}/ \sigma^2$.

\section{Convolutional compressed sensing}\label{sec:CCS}
\par The radar channel $\mathbf{H}\in \mathbb{C}^{M \times N}$ encodes the Doppler shift and AoD information of the targets. Due to the propagation characteristics of the environment at mmWave frequencies, the channel is approximately sparse when expressed in an appropriate basis \cite{HeaGonRan:An-Overview-of-Signal-Processing:16}. For instance, the 2D-DFT basis is often chosen for the sparse representation of $\mathbf{H}$~\cite{JunZhoCha:Beam-codebook-based:09}. Let $\tilde{\bH} \in \mathbb{C}^{M \times N}$ denote the inverse 2D-DFT of $\mathbf{H}$ such that   
\begin{equation}
\mathbf{H}= \mathbf{U}_{M} \tilde{\bH} \mathbf{U}_{N}.
\label{eq:mimoangledom}
\end{equation}
Equivalently, $\tilde{\bH}= \mathbf{U}^{\ast}_{M}\mathbf{H} \mathbf{U}^{\ast}_{N}$ \footnote{The usual convention in radar is that the Doppler-angle channel is the Fourier transform of the time-antenna channel. In this paper, we use the inverse Fourier transform for ease of notation.}.  The matrix $\mathbf{H}$ is the time-antenna domain channel and $\tilde{\bH}$ is called as the Doppler-angle domain channel. The sparse structure in $\tilde{\bH}$ at mmWave allows the use of CS techniques to estimate $\tilde{\bH}$ from fewer radar channel measurements.
\par In this paper, we use a special class of CS called convolutional CS~\cite{li2012convolutional} for sparse radar channel estimation. In this technique, the TX applies fewer circulant shifts of the JCR beamforming vector $\bff_\txm(\delta)$. We use  $c[m]$ to denote the circulant shift used at the TX in the $m^{\mathrm{th}}$ measurement slot. Here, $c[m]$ is an integer in $\{0,1,\cdots, N-1\}$. The beamforming vector applied at the TX is then $\bff_m(\delta) = \bJ_{c[m]} \bff_\txm(\delta)$, where $\bJ_{c[m]} \in \mathbb{C}^{N\times N}$ is the right circulant-delay matrix corresponding to a shift of $c[m]$ units. For example, $\mathbf{J}_1 \in \mathbb{C}^{3\times 3}$ is
\begin{equation}
    \mathbf{J}_1=\left[\begin{array}{ccc}
0 & 1 & 0\\
0 & 0 & 1\\
1 & 0 & 0
\end{array}\right]. 
\end{equation}
In general, $\mathbf{J}_{i}=\mathbf{J}_1 \times \mathbf{J}_1 \times \cdots (i\, \mathrm{times})$. $\mathbf{J}_{0}$ is the $N \times N$ Identity matrix. We substitute the beamforming vector in \eqref{eq:yradFinal} to write 
\begin{align}\label{eq:yG_2}
y_m(\optr,\delta)&= \bee^\rmT_{m,M}  \bU_M \tilde{\bH} \bU_N  \bJ_{c[m]}\bff_\txm(\delta) +  w_m(\optr).
\end{align}
The convolutional structure in the beamforming weights used at the TX allows sparse recovery algorithms to exploit the fast Fourier transform \cite{MyeMezHea:FALP:-Fast-beam:19}.
\par We show that $y_m(\optr,\delta)$ in \eqref{eq:yG_2} can be interpreted as a 2D-DFT measurement of another sparse matrix called the masked Doppler-angle matrix. To define this matrix, we first simplify $\bU_N  \bJ_{c[m]}\bff_\txm(\delta)$ in \eqref{eq:yG_2}. We define a diagonal matrix containing the scaled DFT of $\bff_\txm(\delta)$ on its diagonal as
\begin{equation}\label{eq:Lam}
\boldsymbol{\Lambda}(\delta) = \sqrt{N} \mathrm{diag}(\bU_N \bff_\txm(\delta)). 
\end{equation}
By the property that circulantly shifting a vector modulates the phase of its DFT representation, the DFT of $\bff_m(\delta)= \bJ_{c[m]} \bff_\txm(\delta)$ can be expressed as
\begin{equation} \label{eq:Lambda}
\bU_N  \bJ_{c[m]} \bff_\txm(\delta)  = \boldsymbol{\Lambda}(\delta) \bU_N \bee_{c[m],N}.
\end{equation}
We define $\tilde{{\bZ}}(\delta)=\tilde{\bH} \boldsymbol{\Lambda}(\delta)$ as the masked Doppler-angle matrix. The mask is due to the multiplication effect induced by $\boldsymbol{\Lambda}(\delta)$ on the columns of $\tilde{\bH}$. The matrix $\tilde{{\bZ}}(\delta)$ is sparse as $\tilde{\bH}$ is sparse. As all the diagonal entries in $\boldsymbol{\Lambda}(\delta)$ are non-zero with our design in Algorithm 1, information about all the targets is preserved in $\tilde{{\bZ}}(\delta)$. In this paper, we focus on estimating the sparse matrix $\tilde{{\bZ}}(\delta)$ instead of the sparse $\tilde{\bH}$. Such an approach allows a tractable CS matrix design. Denoting $\mathbf{e}_{c[m],N}$ as a standard basis vector in $\mathbb{C}^N$, the measurement in \eqref{eq:yG_2} can be simplified using  \eqref{eq:Lambda} as
\begin{align}\label{eq:yG}
y_m(\optr,\delta)&= \bee^\rmT_{m,M}  \bU_M \underbrace{\tilde{\bH} \boldsymbol{\Lambda}(\delta)}_{\tilde{{\bZ}}(\delta)}  \bU_N \bee_{c[m],N} +  w_m[\optr] \\
\label{eq:ztilde}
&= \bee^\rmT_{m,M} \underbrace{\bU_M \tilde{{\bZ}}(\delta) \bU_N} _{\mathrm{2D-DFT\,of\, } \tilde{{\bZ}}(\delta)} \bee_{c[m],N} +  w_m[\optr].
\end{align}
We observe from \eqref{eq:ztilde} that the $m^{\mathrm{th}}$ channel measurement with the circulant shift-based training is the $(m, c[m])^{\mathrm{th}}$ entry of the 2D-DFT of $ \tilde{{\bZ}}(\delta)$.
\par
\begin{figure}[h!]
\centering
\includegraphics[width=0.65\columnwidth]{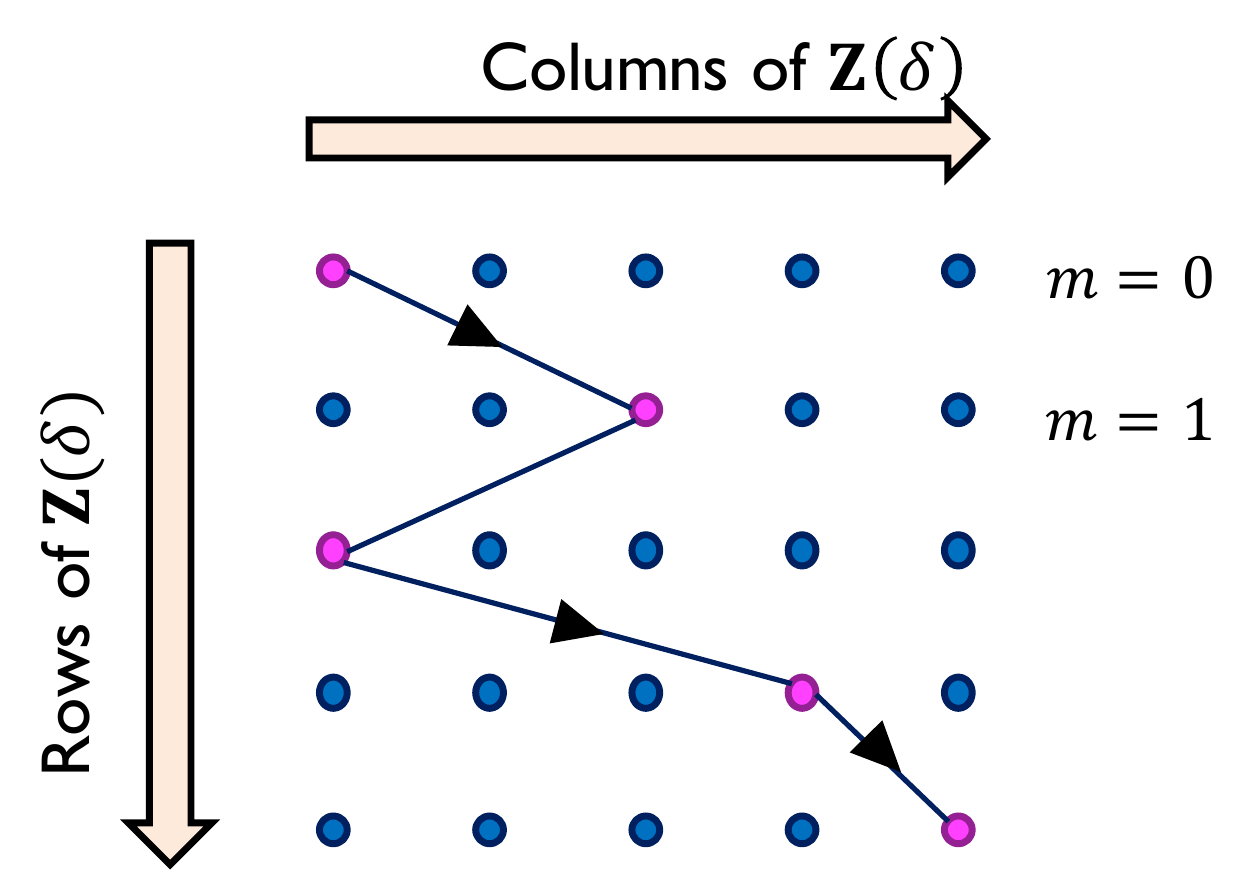}
 \caption{The sampling trajectory traverses through one element in every row of $\bZ (\delta)$ for the combined-waveform beamforming design in the CCS-JCR approach.}
\label{fig:Binarysubsampling}
\end{figure}

 Now, we discuss how sparse recovery of $\tilde{{\bZ}}(\delta)$ is a partial 2D-DFT CS problem and explain the notion of a trajectory. We observe from \eqref{eq:ztilde} that the RX acquires $M$ radar channel measurements when the TX applies $M$ circulant shifts of $\bff_\txm(\delta)$. The measurements are subsamples from an $M \times N$ matrix ${\bZ}(\delta)$ which is defined as
\begin{align}
\label{eq:zmatrix}
{\bZ}(\delta)&=\mathbf{U}_M  \tilde{{\bZ}}(\delta) \mathbf{U}_N.
\end{align}
From \eqref{eq:ztilde} and \eqref{eq:zmatrix}, we notice that our approach obtains the entries of ${\bZ}(\delta)$ at the 2D-coordinates $\{(m,c[m])\}^{M-1}_{m=0}$ on an $M \times N$ grid. A trajectory is defined as a 2D-path on this grid which traverses through ${(m,c[m])}^{M-1}_{m=0}$ in sequence. This set of 2D coordinates is defined as $\Omega$. An example of a trajectory for $M=5$ and $N=5$ is shown in Fig. \ref{fig:Binarysubsampling}. As the goal is to estimate the sparse matrix $\tilde{{\bZ}}(\delta)$ from the subsamples of its 2D-DFT ${\bZ}(\delta)$, the sparse recovery problem is a partial 2D-DFT CS problem \cite{Rau:Compressive-sensing-and-structured:10}. 

\par The reconstruction performance with partial 2D-DFT CS depends on the subsampling trajectory. Prior work has shown that random subsampling trajectories can achieve sparse recovery with partial 2D-DFT CS~\cite{LusDonSan:Compressed-Sensing-MRI:08}. Trajectories that are fully random in the $M \times N$ grid, however, cannot be used in the sparse Doppler-angle estimation problem. We show that the number of feasible 2D-trajectories in our problem is smaller than the trajectories in a typical partial 2D-DFT CS problem. For the measurement slot indexed $m$, the trajectory is at $(m,c[m])$ where the row-coordinate is $m$. For every $m$, the TX can choose $c[m]$ from the $N$ integers in $\{0,1,2,\cdots,N-1\}$. Therefore, the number of feasible trajectories is $N^M$. In a typical partial 2D-DFT CS problem, however, the number of feasible trajectories of length $M$ on an $M\times N$ grid is $MN(MN-1)(MN-2)\cdots (MN-(M-1))$ which is greater than $N^M$. The problem now is to find those trajectories among the $N^M$ feasible candidates that result in better sparse radar channel reconstruction.
\par A reasonable choice for the subsampling trajectory is one that is chosen at random from the $N^M$ feasible candidates. But, is this the best strategy? The answer to this question is not clear at this point. In Section \ref{sec:subOpt}, we propose a novel deterministic subsampling trajectory that achieves better channel reconstruction than a feasible random trajectory.

\section{How to design a good subsampling trajectory?} \label{sec:subOpt}
A good subsampling trajectory is one that results in a CS matrix with the smallest coherence~\cite{CanEldNee:Compressed-sensing-with:11}. This is because a lower coherence results in better sparse recovery~\cite{CanEldNee:Compressed-sensing-with:11}. In this section, we give an explicit form of the CS matrix and derive the desired subsampling trajectory.

We use $\mathbf{A}\in \mathbb{C}^{M \times MN}$ to denote the CS matrix corresponding to the partial 2D-DFT measurement model in \eqref{eq:ztilde}. The $m^{\mathrm{th}}$ row of $\mathbf{A}$ is defined as 
\begin{equation}\label{eq:A}
\mathbf{A}(m,:)= ( \mathbf{e}^\rmT_{c[m],N}\mathbf{U}_N) \otimes ( \mathbf{e}_{m,M}^\rmT \mathbf{U}_M).
\end{equation}
We rewrite \eqref{eq:ztilde} with the radar noise vector $\mathbf{w}[\optr] = [w_0[\optr], w_1[\optr],\cdots, w_{M-1}[\optr]]^\rmT$, the radar channel measurement vector $\by(\optr,\delta) = [y_0(\optr,\delta), y_1(\optr,\delta), \cdots, y_{M-1}(\optr,\delta)]$, and the masked Doppler-angle matrix $\tilde{\bz}(\delta) = \mathrm{vec}(\tilde{\bZ}(\delta))$,  as
\begin{equation}\label{eq:Ay}
\by(\rho,\delta) = \bA \tilde{\bz}(\delta) + \bw[\rho],
\end{equation}
which is the standard CS linear measurement model.

The coherence of the CS matrix $\bA$ in \eqref{eq:A} is defined as~\cite{CanEldNee:Compressed-sensing-with:11}
\begin{equation}
\label{eq:incoh_def1}
\mu= \underset{(i,\ell), i \neq \ell}{\text{max}} \frac{| (\mathbf{A}(:,i))^* \mathbf{A}(:,\ell) |}{\Vert \mathbf{A}(:,i) \Vert \Vert \mathbf{A}(:,\ell) \Vert}.
\end{equation}
Due to the partial 2D-DFT nature of $\mathbf{A}$, the coherence $\mu$ can also be expressed in terms of the point spread function (PSF)~\cite{LusDonPau:Sparse-MRI:-The-application:07}. To explain the PSF, we first define an $M\times N$ binary subsampling matrix $\mathbf{B}$ where
\begin{equation}
\label{eq:binary_mat}
\mathbf{B}(m,n)=\begin{cases}
\begin{array}{c}
1,\,\,\,\,\mathrm{if}\,\,(m,n)\in\Omega\\
0,\,\,\,\,\mathrm{if}\,\,(m,n)\notin\Omega
\end{array}\end{cases}.
\end{equation}
We define $\tilde{\mathbf{B}}$, the 2D-DFT of $\mathbf{B}$, as the PSF. Specifically, $\tilde{\mathbf{B}}=\mathbf{U}_M\mathbf{B} \mathbf{U}_N$. Now, $\mu$ in \eqref{eq:incoh_def1} can also be expressed as~\cite{LusDonPau:Sparse-MRI:-The-application:07,PatEasHea:Compressed-synthetic-aperture:10}
\begin{equation}
\label{eq:incoh_def2}
\mu=\frac{\sqrt{MN}}{M}\underset{(p,\fftn) \neq (0,0)}{\text{max}}|\tilde{\mathbf{B}}(p,\fftn)|.
\end{equation}
The focus of this section is to construct the subsampling set $\Omega=\{(m,c[m])\}_{m=0}^{M-1}$ that results in the smallest $\mu$ under the sampling constraints in our problem.
\par Now, we discuss the structure of the PSF $\tilde{\mathbf{B}}$ under the sampling constraints. We observe from \eqref{eq:binary_mat} that the $m^{\mathrm{th}}$ row of $\mathbf{B}$ has a single one in the $c[m]^{\mathrm{th}}$ column and has zeros at the other locations for $\Omega=\{(m,c[m])\}^{M-1}_{m=0}$. The $M \times N$ binary subsampling matrix is 
\begin{equation}
\label{eq:B_explicit}
\mathbf{B}=\begin{pmatrix}
\begin{array}{c}
\mathbf{e}^T_{c[0],N}\\
\mathbf{e}^T_{c[1],N}\\
\vdots\\
\mathbf{e}^T_{c[M-1],N}
\end{array}
\end{pmatrix}.
\end{equation} 
We define $\omega=\mathrm{exp}(-\mathrm{j}2\pi/N)$ and compute the PSF $\tilde{\mathbf{B}}$ from $\mathbf{B}$ in \eqref{eq:B_explicit}. First, we find the $N$-point DFT of every row in $\mathbf{B}$. Since $\mathbf{e}^T_{c[m],N} \mathbf{U}_N=[1, \omega^{c[m]}, \omega^{2c[m]}, \cdots, \omega^{(N-1)c[m]}]/\sqrt{N}$, we can write 
 \begin{equation}
 \label{eq:BUN}
 \mathbf{B}\mathbf{U}_N= \begin{small} \frac{1}{\sqrt{N}}\left[\begin{array}{ccccc}
1 & \omega^{c[0]} & \omega^{2c[0]} & \cdots & \omega^{(N-1)c[0]}\\
1 & \omega^{c[1]} & \omega^{2c[1]} & \cdots & \omega^{(N-1)c[1]}\\
\vdots & \vdots & \vdots &  & \vdots\\
1 & \omega^{c[M-1]} & \omega^{2c[M-1]} & \cdots & \omega^{(N-1)c[M-1]}
\end{array}\!\!\right].
\end{small}
 \end{equation}
To express \eqref{eq:BUN} in compact form, we define $\mathbf{g} \in \mathbb{C}^M$ as 
\begin{equation}
\label{eq:gdef}
\mathbf{g}=[\omega^{c[0]},\omega^{c[1]}, \cdots,\omega^{c[M-1]}]^T.
\end{equation}
Then, 
\begin{equation}
\mathbf{B}\mathbf{U}_N=[\mathbf{g}^0,\mathbf{g}^1,\mathbf{g}^2,\cdots,\mathbf{g}^{N-1}]/\sqrt{N}. 
\end{equation}
Note that $\mathbf{g}^0 = \mathbf{1}$, where $\mathbf{1}$ is an all-ones vector of length $M$. Now, the PSF $\tilde{\mathbf{B}}=\mathbf{U}_M \mathbf{B} \mathbf{U}_N$ is obtained by taking the $M$-point DFT of every column in $\mathbf{B}\mathbf{U}_N$, i.e., 
\begin{equation}
\label{eq:PSFexplicit}
\tilde{\mathbf{B}}= [\mathbf{U}_M \mathbf{g}^0,\mathbf{U}_M \mathbf{g}^1, \mathbf{U}_M \mathbf{g}^2,\cdots,\mathbf{U}_M \mathbf{g}^{N-1}]/ \sqrt{N}.
\end{equation}
The problem now is to find a vector $\mathbf{g}$ of the form in \eqref{eq:gdef} such that the PSF in \eqref{eq:PSFexplicit} achieves the smallest coherence.
\par We now examine the entries of the PSF. The first column of $\tilde{\mathbf{B}}$ in \eqref{eq:PSFexplicit} is the DFT of $ \mathbf{1}/\sqrt{N}$, which is the $M$ length vector $\sqrt{M/N}\mathbf{e}_{0,M}$. As all the entries in the first column other than $\tilde{\mathbf{B}}(0,0)$ are $0$, this column does not impact $\mu$ defined in \eqref{eq:incoh_def2}. The other columns of $\tilde{\mathbf{B}}$ which have the form $\mathbf{U}_M \mathbf{g}^\fftn$ for $\fftn\neq 0$ determine $\mu$. To achieve the smallest coherence, the largest entry of $|\mathbf{U}_M \mathbf{g}^\fftn|$ must be minimized for every $\fftn\in  \{1,2,3,...,N-1\}$. As $\Vert \mathbf{g}^\fftn \Vert =\sqrt{M}$, it follows that $\Vert \mathbf{U}_M \mathbf{g}^\fftn \Vert =\sqrt{M}$ for every $\fftn$. Under this norm constraint, the largest entry of $|\mathbf{U}_M \mathbf{g}^\fftn|$ can be no smaller than $1$. Therefore, we seek a $\mathbf{g}$ such that 
\begin{equation}
\label{eq:unimod_condn}
|\mathbf{U}_M \mathbf{g}^\fftn |=\mathbf{1}, \, \forall \fftn \in \{1,2,3,...,N-1\},
\end{equation}
i.e., the DFT of every $\fftn^{\mathrm{th}}$ power of $\mathbf{g}$ must have a constant magnitude for $1 \leq \fftn \leq N-1$. Furthermore, $\mathbf{g}$ must be expressible in the form of \eqref{eq:gdef}. Can we find such a $\mathbf{g}$?
\par We discuss why the Zadoff-Chu (ZC) sequence is a reasonable choice for the vector $\mathbf{g}$. We use $\bzc \in \mathbb{C}^M$ to denote a ZC sequence of root $u$ and length $M$. Here, $u$ is co-prime with $M$. The $m^{\mathrm{th}}$ entry of $\bzc$ is~\cite{Chu:Polyphase-codes-with:72}
\begin{equation}
\label{eq:zcdefn}
\xi[m]=
\begin{cases}
\begin{array}{c}
\mathrm{exp}\left(-\mathrm{j} \frac{\pi u m\left(m+1\right)}{M}\right),\,\,\,\,\,\mathrm{if}\,M\,\mathrm{is\,odd}\\
\mathrm{exp}\left(-\mathrm{j} \frac{\pi u m^{2}}{M}\right),\,\,\,\,\,\,\,\,\,\,\,\,\,\,\,\,\mathrm{if}\,M\,\mathrm{is\,even}
\end{array}\end{cases}.
\end{equation}
An interesting property is that the DFT of a ZC sequence has a constant magnitude, i.e., $|\mathbf{U}_M\bzc|=\mathbf{1}$~\cite{Luk:Sequences-and-arrays-with:88}. When $\mathbf{g}$ is set to $\bzc$, we observe that the unimodular DFT condition in \eqref{eq:unimod_condn} holds when $\fftn=1$. Now, we notice from \eqref{eq:zcdefn} that $\mathbf{g}^\fftn=\bzc^\fftn$ has the same structure as $\bzc$, but with root $\fftn u$ instead of $u$. When $\fftn u$ is coprime with $M$, $\mathbf{g}^\fftn$ is a ZC sequence that satisfies $|\mathbf{U}_M \mathbf{g}^\fftn|=\mathbf{1}$. Therefore, the condition in \eqref{eq:unimod_condn} is met when $\mathbf{g}$ is a ZC sequence of root $u$ and when $\fftn u$ is coprime with $M$ for $\fftn\in \{1,2,3, \cdots, N-1\}$. One way to ensure that the co-prime condition is met is to set $u=1$, and $M$ to a prime number that is not smaller than $N$. As the prime number $M$ is also an odd number, the ZC sequence $\mathbf{g}$ is such that 
\begin{equation}
\label{eq:g_zc}
g[m]=\mathrm{exp}\left(-\mathrm{j} \frac{\pi m\left(m+1\right)}{M}\right).
\end{equation}
The subsampling design problem is solved when such a $\mathbf{g}$ can also be expressed in the form of \eqref{eq:gdef}.

\begin{figure}[!t]
\begin{minipage}[b]{\linewidth}
  \centering
  \centerline{ \includegraphics[clip,width=0.8\columnwidth]{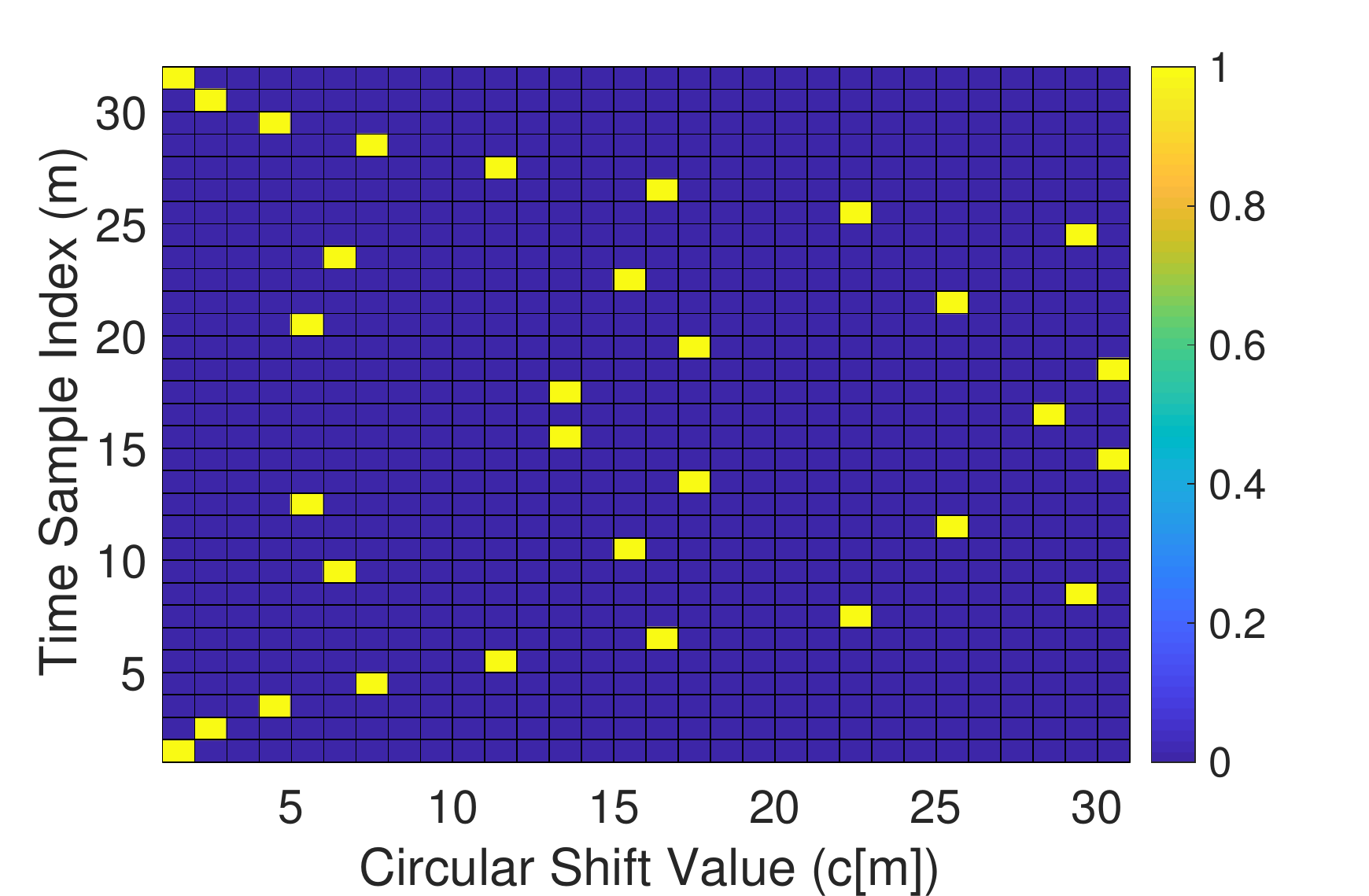}}
  \centerline{(a) Binary subsampling matrix}\medskip
\end{minipage}
\begin{minipage}[b]{\linewidth}
  \centering
  \centerline{\includegraphics[clip,width=0.8\columnwidth]{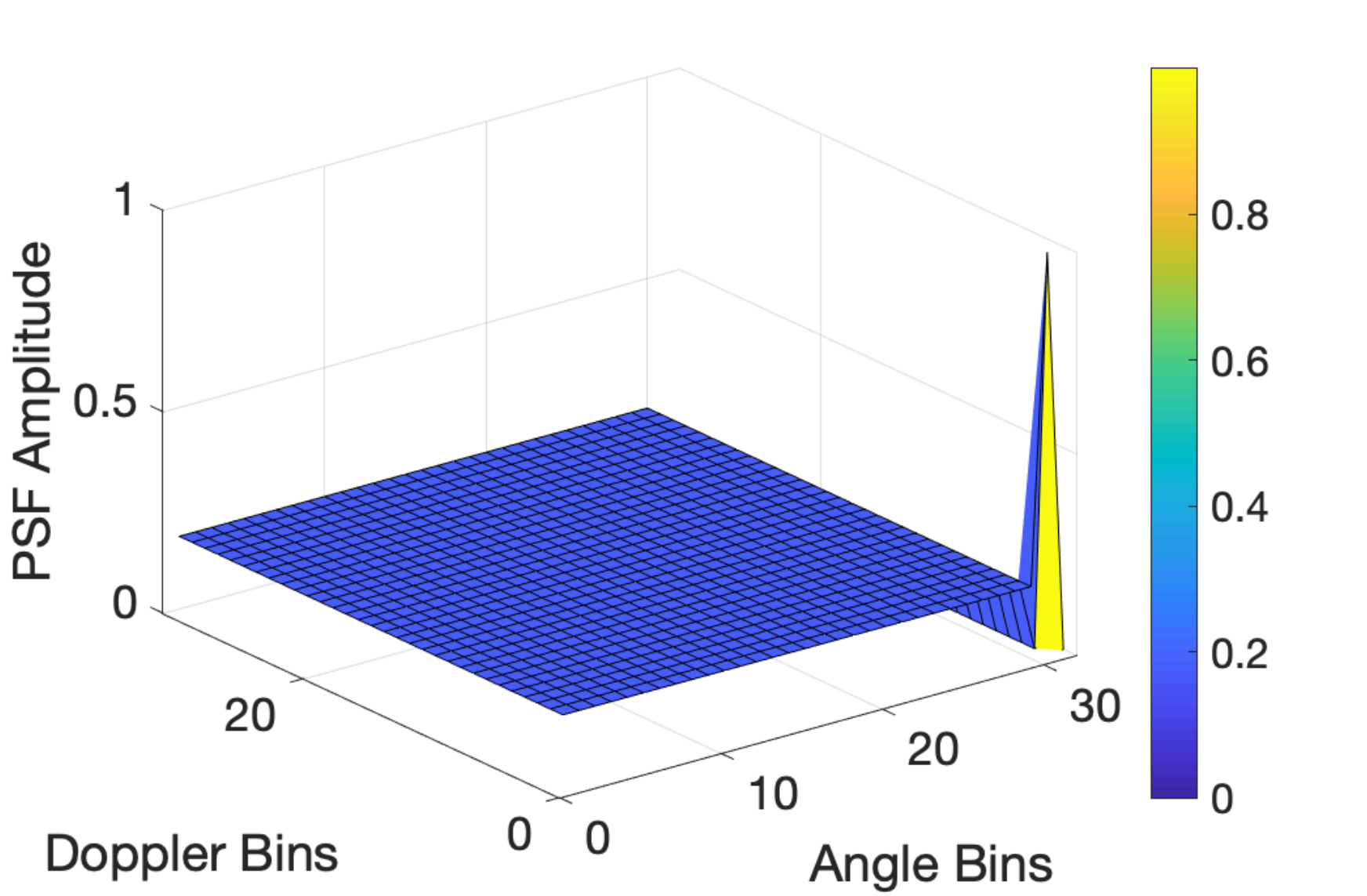}}
  \centerline{(b)  PSF of the binary subsampling matrix}\medskip
\end{minipage}
\caption{An example of the optimized binary sampling matrix and its corresponding PSF for $M=N=31$. The PSF has constant amplitude of $1/\sqrt{31}$ in all the columns except the first one. The first column in the PSF matrix is $\bee_{0,31}$.}
\label{fig:TrajResult}
\vspace{-1.2em}
\end{figure}

\par Now, we find conditions on $M$ and $N$ by setting $\mathbf{g}$ in \eqref{eq:g_zc} to the one in \eqref{eq:gdef}. We observe that the phase of the $g[m]$ in \eqref{eq:gdef} is $-2\pi c[m]/N$, while the phase of $g[m]$ in \eqref{eq:g_zc} is $-\pi m(m+1)/M$. The two vectors in \eqref{eq:gdef} and \eqref{eq:g_zc} are equal when 
\begin{equation}
\frac{2 \pi c[m]}{N}=\frac{\pi m\left(m+1\right)}{M} +2 \pi i_m, \, \forall m\in \{0,1,2,\cdots,M-1\}, 
\end{equation}
for some integers $\{i_m\}^{M-1}_{m=0}$. Equivalently, 
\begin{equation}
\label{eq:circhift_pre}
c[m]=\frac{m \left(m+1\right)N}{2M} + N i_m.
\end{equation}
Note that $c[m]$ has to be an integer in $\{0,1,2,\cdots, N-1\}$ as it models the circulant shift applied at the TX for the $m^{\mathrm{th}}$ channel measurement. It can be observed that a circulant shift of $N$ is equivalent to a zero circulant shift. As $Ni_m$ is an integer multiple of $N$, it does not contribute to $c[m]$ in \eqref{eq:circhift_pre}. Therefore, for $c[m]$ to be an integer, we set $M=N$ in \eqref{eq:circhift_pre} to obtain 
\begin{equation}
\label{eq:circhift_fin}
c[m]=\left[\frac{m \left(m+1\right)}{2}\right]_{\mathrm{mod}\, N}.
\end{equation}
Note that the subsampling coordinates in partial 2D-DFT CS are $(m,c[m])^{M-1}_{m=0}$. Such a subsampling technique achieves the smallest coherence under the constraints in our problem. We would like to mention that our result is valid when $M$ is prime and when $M=N$. Optimizing subsampling for other settings is an interesting research direction.
\par We now discuss the practical aspects of the designed subsampling technique. An example of the binary subsampling matrix is shown in Fig.~\ref{fig:TrajResult}(a). We plot the PSF corresponding to this matrix in Fig. \ref{fig:TrajResult}(b). 
As $|\mathbf{U}_M \mathbf{g}^\fftn|=\mathbf{1}$ with our construction, we observe from \eqref{eq:PSFexplicit} and \eqref{eq:incoh_def2} that $\mu=1/ \sqrt{M}$. This is also $1/\sqrt{N}$ as $M=N$ with our design. Note that the subsampling ratio in our setting is $M/(MN)$ which is $1/N$. Prior work has shown that the standard OMP algorithm can recover upto $0.5 \left(1+ 1/{\mu}\right)$ sparse coefficients when the CS matrix has a coherence of $\mu$~\cite{CaiWan:Orthogonal-matching-pursuit:11}. Therefore, the proposed partial 2D-DFT CS technique that acquires subsamples defined by \eqref{eq:circhift_fin} can recover
\begin{equation}\label{eq:csK}
K<\frac{1}{2}\left(1+\sqrt{N} \right)
\end{equation}
targets in the Doppler-angle space. In the simulations section, we will show that the OCCS-JCR with the designed subsampling trajectory achieves better radar detection performance than the RCCS-JCR with random subsampling trajectory.

\section{Adaptive JCR design}\label{sec:AdapJCR}
In this section, we quantity the JCR trade-off between radar and communication for an adaptive combined waveform-beamfoming design using MSE-based metrics. We first describe the NMSE metric for radar, followed by the DMSE metric for communication. Additionally, we also present a MSE-based adaptive combined waveform-beamforming design for the mmWave automotive JCR.

\subsection{Radar performance metric}
We use the NMSE metric to evaluate the performance of the CS-based radar channel estimation algorithm for our adaptive JCR design. Without loss of generality, we assume the average target channel power as one. The NMSE metric for a true Doppler-angle domain radar channel $\tilde{\bh} = \mathrm{vec}( \tilde{\bH}) $ and the estimated Doppler-angle domain radar channel $\tilde{\bh}_\mathrm{est}$ is defined as
\begin{equation}~\label{eq:NMSE}
\nmse(\rho,\delta) \triangleq \frac{1}{K}\e{{\vert \vert \tilde{\bh} -\tilde{\bh}_\mathrm{est}(\optr,\delta) \vert \vert^2}},
\end{equation}
where $K$ is the number of targets.

\begin{figure}[!t]
\centering
\includegraphics[width=0.75\columnwidth]{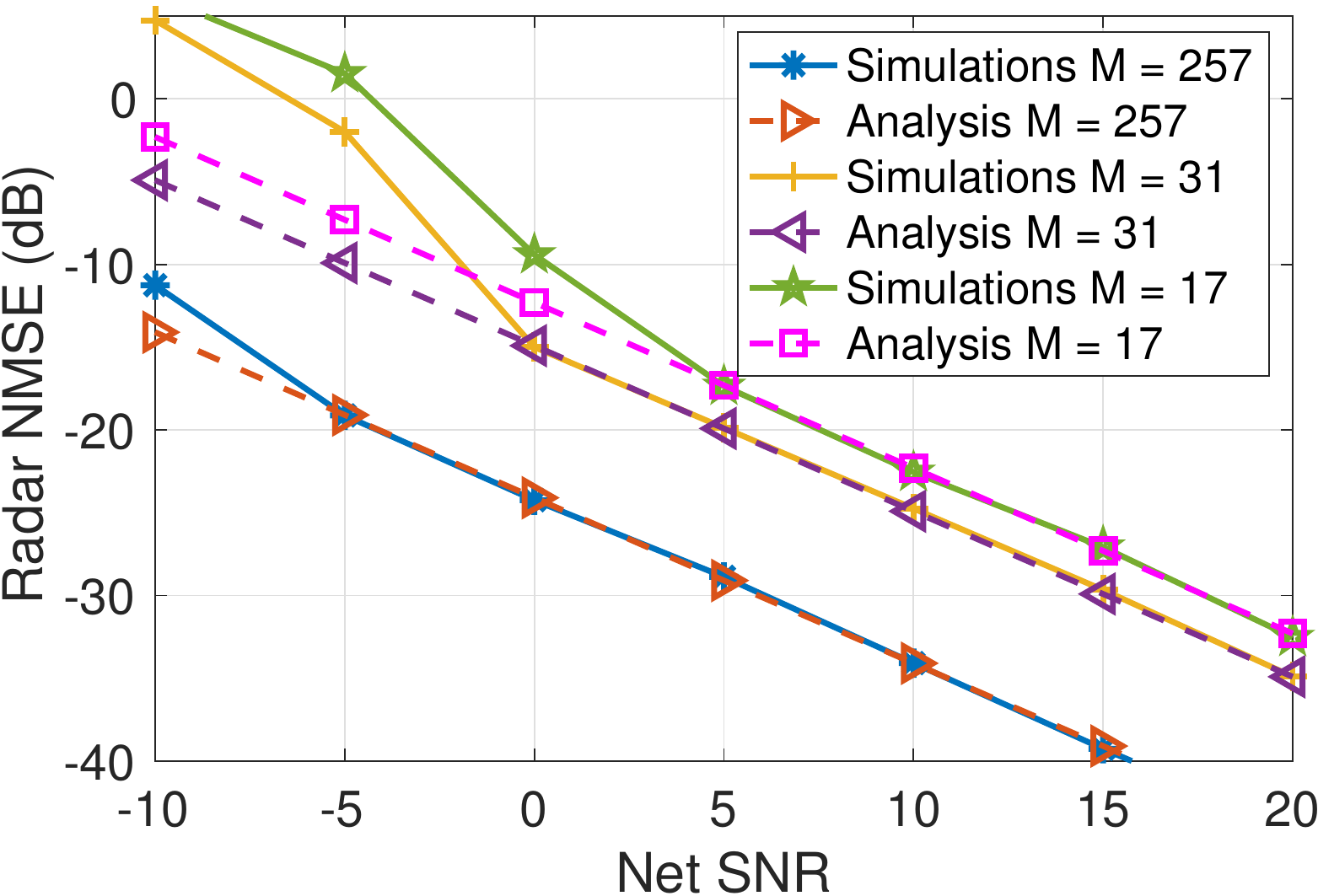}
 \caption{Comparison between the radar NMSE obtained using analysis in~\eqref{eq:NMSEapprox} and using simulations with varying net SNR for a single-target scenario. The analytical NMSE closely matches with those of simulations above a certain SNR value that decreases with increasing $M$. }
\label{fig:NMSEanalysis}
\vspace{-1.2em}
\end{figure}

The NMSE for estimating the masked Doppler-angle channel vector $\tilde{\bz}(\delta)$ in \eqref{eq:Ay} corresponding to the dominant channel taps using the optimized sampling trajectory in~\eqref{eq:circhift_fin} and the OMP estimation can be approximated similar to the NMSE derivation in~\cite{TanWuHer:Performance-Analysis-of-OMP-Based:18}. From \eqref{eq:DFTmagf} and \eqref{eq:Lam}, we see that the amplitude of all the diagonal elements in $\boldsymbol{\Lambda}(\delta)$ is $\sqrt{N \delta_\mr}$ except the first one. Therefore, denoting $\bA_\mathrm{S}$ as the matrix formed by the support columns (defined as the final set of dominant columns chosen from the dictionary), the pseudoinverse of matrix $\bA_\rmS$ as $\bA_\rmS^+ = (\bA_\rmS^*\bA_\rmS)^{-1}\bA_\rmS^*$, the net radar SNR as $\zeta_\mathrm{net}(\rho,\delta) = N \delta_\mr \zeta_\rmp[\optr]$, the NMSE for estimating the Doppler-angle channel vector $\tilde{\bh}$ corresponding to the MRR/SRR targets can be approximated similar to~\cite{TanWuHer:Performance-Analysis-of-OMP-Based:18} as
\begin{equation} \label{eq:NMSEapprox}
\nmse(\rho,\delta) \approx \frac{1}{K \zeta_\mathrm{net}(\rho,\delta)} \Tr{\bA^+_{\rmS}(\bA^+_{\rmS})^*},
\end{equation}
when the probability of success of OMP exceeds a certain threshold.
We see from \eqref{eq:NMSEapprox} that the NMSE approximation is inversely proportion to the net radar SNR, which is a linear function of $\rho$ and an approximate linear function of $\delta$, where the approximation to the linearity is because of the use of phased array architecture and the GS algorithm.

Fig.~\ref{fig:NMSEanalysis} shows the comparison between the NMSE obtained using~\eqref{eq:NMSEapprox} and the NMSE estimated using simulations with varying net SNR $\zeta_\mathrm{net}(\rho,\delta)$ for a single-target scenario. The analytical NMSE closely matches above a certain net SNR value. The net SNR value where the analytical and simulations NMSE closely matches decreases with the increase in $M$. In Section~\ref{sec:Results}, we will further explore the relation between the radar NMSE and radar SNR for different multi-target scenarios.

\subsection{Communication performance metric}
Assuming $s_{m,\ell}[\optr]$ is distributed as $\mathcal{N}(0,1)$, the maximum achievable communication spectral efficiency, $r$, for a JCR system with $\optr= 0$ and $T_{\mathrm{IFS}}=0$ is given by
\vspace{-0.4em}
\begin{equation} \label{eq:cap1}
r(\delta) =   \log_2 \left(1 + \delta \zeta_\mcom \right) ,
\end{equation}
where $\zeta_\mcom = \Es \vert h_\mcom \vert^2 N^2 /\sigma^2_\mcom$. The channel capacity in bits per second (bps) is given as $C(\delta) = Wr(\delta)$, and the communication minimum MSE (MMSE) per symbol is given as~\cite[Ch.~7]{HeaLoz:Foundations-of-MIMO-Communication:19}
\begin{equation} \label{eq:mmse}
\mathrm{MMSE}(\delta) = \frac{1}{1+\delta \zeta_\mcom} = 2^{-r(\delta)},
\end{equation} 
and $\log_2 \mathrm{MMSE}(\delta) = -r(\delta) $ is a logarithmic function of $\delta$.

When $\optr>0$ or and $T_{\mathrm{IFS}}>0$, the effective maximum achievable communication spectral efficiency, $r_\eff$, decreases by a factor of $\alpha[\rho]$
\begin{equation}
\alpha[\optr] = 1 - \frac{M (\optr\Ts+ T_\mathrm{IFS})}{T},
\end{equation}
and we define the effective communication spectral efficiency as~\cite[Ch.~7]{HeaLoz:Foundations-of-MIMO-Communication:19}
\begin{align} \label{eq:capAlpha}
r_\eff(\optr,\delta)  &= \alpha[\optr]  \log_2 \left(1 + \delta \zeta_\mcom \right)  \text{\, bits/s/Hz} \\&=   r(\delta)^ {\alpha[\optr]}\text{\, bits/s/Hz}.
\end{align}
The effective channel capacity in bps is given as $C_\eff(\optr,\delta) = Wr_\eff(\optr,\delta)$. We see from $\eqref{eq:capAlpha}$ that the effective communication spectral efficiency is linearly related to $\rho$, whereas it is logarithmically related to $\delta$.

\subsection{JCR performance metric}

The performance metrics of radar and communication are dependent on $\optr$ and $\delta$, as can be seen from (\ref{eq:NMSE}), \eqref{eq:NMSEapprox}, and (\ref{eq:capAlpha}). With an increase in $\delta$ and a decrease in $\rho$, the information rate improves, whereas the NMSE for radar channel estimation degrades. Therefore, we focus on optimizing $(\rho,\delta)$ for the adaptive mmWave automotive JCR combined waveform-beamforming design. This requires the use of a comparable metric to accurately quantify both radar and communication system performances.  

To use an effective scalar communication metric that parallels the concept of the radar NMSE for JCR waveform design optimization, we use an effective distortion MSE communication metric analogous to the distortion metric in the rate-distortion theory~\cite[Ch. 10]{CovTho:Elements-of-information-theory:12}, which is defined as~\cite{KumVorHea:Adaptive-Virtual-Waveform:20}
\begin{equation} \label{eq:DMSEeff}
\dmseeff(\rho,\delta) = 2^{-r_\eff(\rho,\delta)} = \left( \mathrm{MMSE}(\delta) \right)^ {\alpha[\rho]}.
\end{equation} 
According to \eqref{eq:mmse} and \eqref{eq:DMSEeff}, each bit of description reduces the communication distortion MSE by a factor of 2. This implies that as the effective spectral efficiency decreases by a factor of $\alpha[\rho]$, the effective average DMSE increases exponentially by the same factor. Since there is a simple one-to-one relation between effective spectral efficiency and effective DMSE, and the expressions \eqref{eq:mmse} and \eqref{eq:DMSEeff} are analogous to the relation between mean squared-error distortion and rate in the rate distortion theory\cite[Ch. 10]{CovTho:Elements-of-information-theory:12}, it is easy to use and understand. Additionally, this metric is easily extendable to other automotive JCR scenarios, such as the multi-target situation~\cite{KumVorHea:Adaptive-Virtual-Waveform:20}, unlike the radar estimation rate metric in~\cite{Bli:Cooperative-radar-and-communications:14}. 

Since the communication DMMSE and the radar CRB values are usually substantially different, the log-scale is used to achieve proportional fairness similar to the problem of resource allocation in multi-user communication \cite[Ch. 7]{HeaLoz:Foundations-of-MIMO-Communication:19}. The performance trade-off between communication and radar can then be quantified in terms of the following scalar quantities: ${\log (\dmseeff})$ and ${\log(\nmse})$.

\subsection{Weighted-average optimization-based JCR design} \label{sec:JCRdesign}

Now, we formulate an adaptive JCR combined waveform-beamforning design to optimize the preamble block count $\rho$ and the fraction of communication TX gain $\delta$. The JCR performance optimization problem is a multi-objective (two-objective) problem of simultaneously optimizing both the radar performance, in terms of, for example, minimizing the radar NMSE, and the communication performance, in terms of minimizing the effective communication DMSE. We can see from (\ref{eq:DMSEeff}) that the communication DMSE metric denoted as $ \log \dmseeff$ is linear with respect to optimization variables $\rho$ and is logarithmic with the optimization variable $\delta$. The radar NMSE metric denoted as ${\log \nmse}$, however, can be non-convex sometimes with respect to the optimization variables, as illustrated in Fig.~\ref{fig:NMSEanalysis} and later in Section~\ref{sec:Results}. Therefore, the region of achievable JCR objective values with the radar NMSE and communication DMSE pairs corresponding to the feasible values of $\rho$ and $\delta$ can be non-convex. Then, the optimal JCR performance is achieved by using the Pareto set of the minimum convex set (termed the convex hull) of the feasible non-convex JCR achievable objective values region, thereby enhancing at least radar NMSE metric without degrading the communication DMSE metric, similar to multi-user communication rate optimization~\cite[Ch. 15]{CovTho:Elements-of-information-theory:12}. Additionally, the convex solution is achievable by using time-sharing or probabilistic occurrence techniques on the extreme points of the convex hull~\cite{brehmer2012utility}.


The scalarization approach is known to achieve a Pareto optimal point for multiple convex objectives \cite[Ch. 4]{BoyVan:Convex-optimization:04}. Therefore, the JCR performance optimization can be formulated as the weighted average of a convex hull of communication and radar MSE-based performance metrics. We denote the scalar communication DMSE metric as $\varphi_\mcom(\dmseeff) \triangleq {{ \log \dmseeff}}$ and the scalar radar NMSE metric as $\varphi_\mr(\nmse) \triangleq \cvx{{\log \nmse}}$, which incorporates the convex hull operation with respect to the optimization variables. For a given TX precoder codebook $\mathcal{F}(\delta)$ and a maximum preamble building block count of $P_\mathrm{max}$, the JCR performance optimization problem can be formulated as 
\begin{eqnarray} \label{eq:weighted}
&\underset{\rho,\delta}{\text{minimize}}&\;\omega_\mr {\varphi}_\mrad(\nmse)    + \omega_{\comm}  {\varphi}_\comm(\dmseeff)   \nonumber \\
&{\text{subject to}}& \{ T, K, d \} =  \mathrm{constants,} \nonumber\\
&\text{}& \{ \mathbf{f}_m(\delta) \}_{m=0}^{M-1} \in \mathcal{F}(\delta) \nonumber\\
&\text{}&  0 \leq \rho \leq P_\mathrm{max}, \text{ } \rho \in \mathbb{Z}  \nonumber\\
&\text{}&  0 \leq \delta \leq 1, \text{ } \delta \in \mathbb{R} ,
\end{eqnarray}
where $\omega_\mr \geq 0 $ and $\omega_{\comm} \geq 0 $ are the normalizing and weighting factors assigning the priorities for radar and communication tasks, respectively. Note that the weights can be adjusted adaptively with respect to the requirements imposed by different scenarios, such as varying radar SNR. Alternatively, the problem in \eqref{eq:weighted} can be modified as minimization of one of the objectives with second as a constraint that would guarantee an acceptable performance for one of the tasks.

\section{Numerical results} \label{sec:Results}
In this section, the numerical results of the proposed adaptive combined waveform-beamforming design for mmWave automotive JCR are presented. First, we evaluate and compare the radar NMSE performance of OCCS-JCR, RCCS-JCR, and RS-JCR with varying distance, target counts, and number of frames/antenna elements. Then, we study the optimal JCR designs for the weighted average based formulation. For illustration purposes, we consider simulation parameters based on the IEEE 802.11ad standard~\cite{ieee2012wireless,KumChoGon:IEEE-802.11ad-Based-Radar::18} in application to automotive scenarios~\cite{PatTorWan:Automotive-Radars:-A-review:17,HasTopSch:Millimeter-wave-technology-for-automotive:12}. The TX and RX antenna arrays are considered to be uniform linear arrays with 17, 31, and 257 elements. We assume 180$^\circ$ FoV, the recipient vehicle distance $d_\comm = 100$~m, the preamble building block size of 512 symbols, and a coherent processing interval of 5~ms. To estimate the sparse radar channel, we employ the OMP algorithm that exploits the partial 2D-DFT-based structure of the measurement model~\cite{MyeMezHea:FALP:-Fast-beam:19}. Such an algorithm exploits the fast Fourier transform and has a lower complexity than the standard counterparts.

\subsection{Radar performance}
In this subsection, we investigate the radar NMSE performance of our proposed CCS-JCR design. We also compare our proposed OCCS-JCR design with optimized subsampling trajectory developed in Section~\ref{sec:subOpt} versus the RCCS-JCR technique with random sampling. Additionally, we compare the performance of OCCS-JCR and RCCS-JCR designs against the RS-JCR technique.

\begin{figure}[!htb]
\begin{minipage}[b]{\linewidth}
  \centering
  \centerline{ \includegraphics[clip,width=\columnwidth]{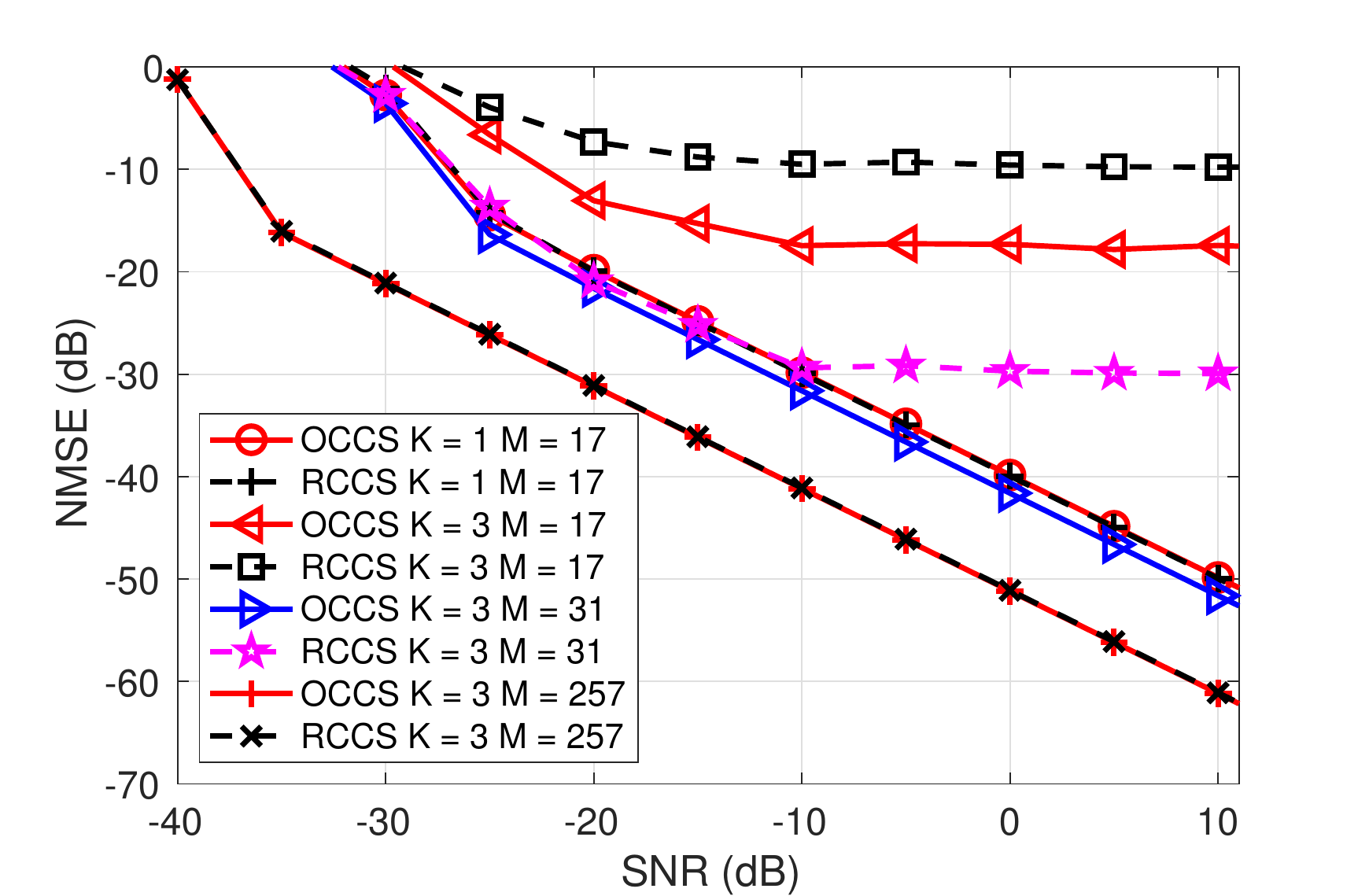}}
  \centerline{(a) NMSE versus SNR}\medskip
\end{minipage}
\begin{minipage}[b]{\linewidth}
  \centering
  \centerline{\includegraphics[clip,width=\columnwidth]{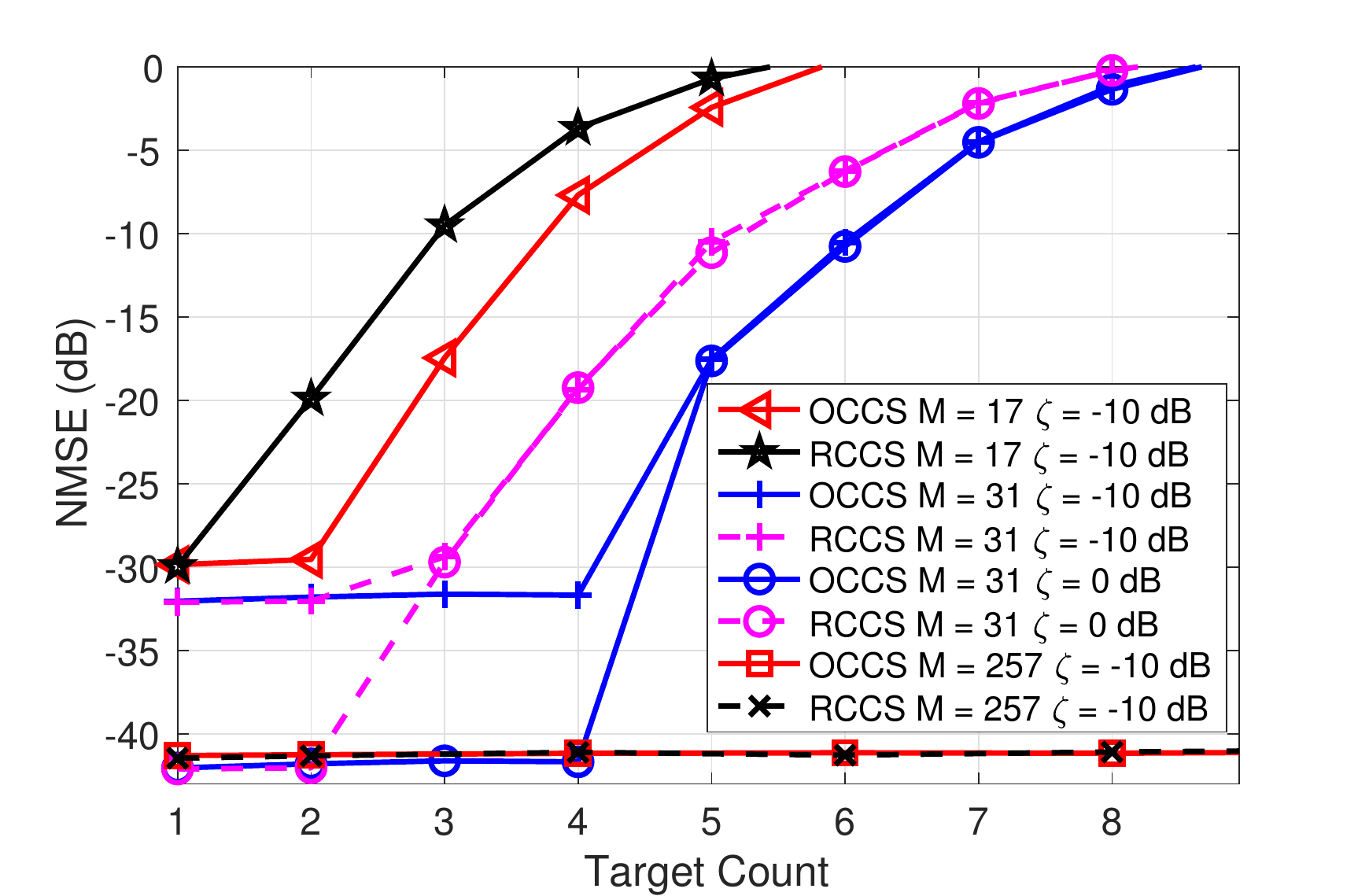}}
  \centerline{(b) NMSE versus target count}\medskip
\end{minipage}
\caption{Comparison between OCCS-JCR and RCCS-JCR for different $M$, $N$, SNR ($\zeta$), and $K$ at $\delta = 0.5$ and 1024 training symbols per frame with $\rho = 2$. The OCCS-JCR design with optimized circulant shifts performed better than the RCCS-JCR design, especially at high SNR and target counts. }
\label{fig:CompareRCCS}
\vspace{-1.2em}
\end{figure}

Figs.~\ref{fig:CompareRCCS}(a) and (b) show the radar performance of OCCS-JCR and RCCS-JCR for different SNR $\zeta$, number of targets $K$, frame counts $M$, and number of TX antenna elements $N$ at $\delta = 0.5$ and $\rho = 2$. In Fig.~\ref{fig:CompareRCCS}(a) the radar NMSE generally decreases with SNR linearly in the logarithmic scale, as also seen from \eqref{eq:NMSEapprox}. At high SNR and with a large number of targets, however, we see the saturation effect where the radar NMSE remains constant. The non-linearity of radar NMSE with SNR is also observed at low SNR. In Fig.~\ref{fig:CompareRCCS}(b), the radar NMSE remains almost the same with increasing number of targets till a critical $K$. The number of targets that satisfies the constraint in \eqref{eq:csK}, do not suffer from the saturation effect at high SNR.  After crossing the critical $K$, the radar NMSE degrades rapidly, and the critical $\zeta$ and $K$, where the saturation happens, increase with $M$. 

From Figs.~\ref{fig:CompareRCCS}(a) and (b), we also see that our proposed optimized CCS-JCR always performs the best. The performance gap between the OCCS-JCR and RCCS-JCR grows with increasing SNR and target count. The critical $K$, where the saturation occurs, is larger in OCCS-JCR than the RCCS-JCR. The performance gap, however, reduces with increasing $M=N$. This reduction with $M$ is because a random trajectory-based sampling matrix results in a small coherence for a large sample space~\cite{Rau:Compressive-sensing-and-structured:10}. As a result, RCCS-JCR approaches the performance of OCCS-JCR for a large $M$.

\begin{figure}[!t]
\centering
\includegraphics[width=\columnwidth]{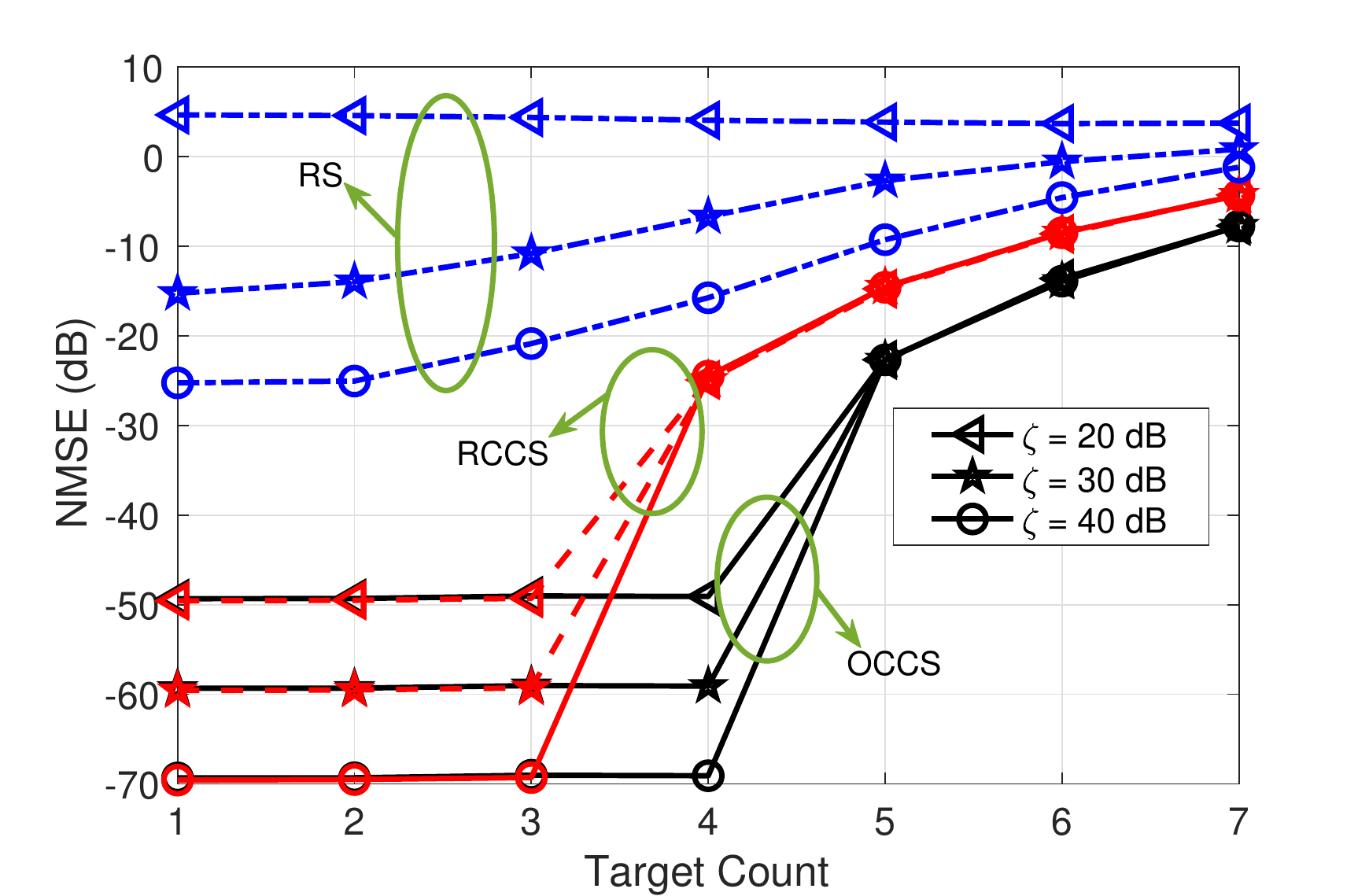}
 \caption{Comparison between OCCS-JCR,  RCCS-JCR, and RS-JCR for different SNRs and varying target counts at $M$ = 31, $\delta = 30/31$, and 1024 training symbols per frame. The OCCS-JCR technique performs the best, followed by RCCS-JCR, and RS-JCR performs the worst. }
\label{fig:CompareRS}
\vspace{-1.2em}
\end{figure}
Fig.~\ref{fig:CompareRS} shows the comparison between the RS-JCR, RCCS-JCR, and OCCS-JCR for different SNRs and varying target counts at $M$ = 31, $\delta = 30/31$, and $\rho=2$. The OCCS-JCR technique performs the best, followed by RCCS-JCR, and RS-JCR performs the worst. The performance gap between RS-JCR and RCCS-JCR is much larger than the gap between RCCS-JCR and OCCS-JCR at small number of targets. The performance gaps between different JCR approaches start decreasing at large $K$ and high SNR. The poor performance of random switching is observed because the CS matrix in this approach has a low coherence in CS and suffer from SNR loss under the per-antenna power constraint. Furthermore, the RCCS-JCR technique suffers from a low SNR under the per-antenna power constraint~\cite{KumMyeVor:A-Combined-Waveform-Beamforming-Design:19}. Since our proposed OCCS-JCR technique performs the best, we will use this approach for the numerical analysis on the optimal JCR design.

\subsection{Optimal JCR designs}
In this subsection, we explore the OCCS-JCR performance trade-off curve between the radar NMSE and the communication NMSE with respect to $\rho$ and $\delta$. Additionally, we investigate the optimal solutions for the weighted average optimization-based JCR design for different SNRs, target counts, and number of frames/antenna elements. We vary the communication weighing $0 \leq \omega_\mcom \leq 1$. The preamble building block length for channel estimation is considered as 512, similar to the IEEE 802.11ad standard. The maximum preamble length is considered as the maximum frame length for $M=257$, which leads to $P_\mathrm{max} = 53$. In our optimization, we do not consider $\delta =0$ because it is unfavorable for both vehicular communication as well as LRR sensing.  

\begin{figure}[!t]
\centering
\includegraphics[width=\columnwidth]{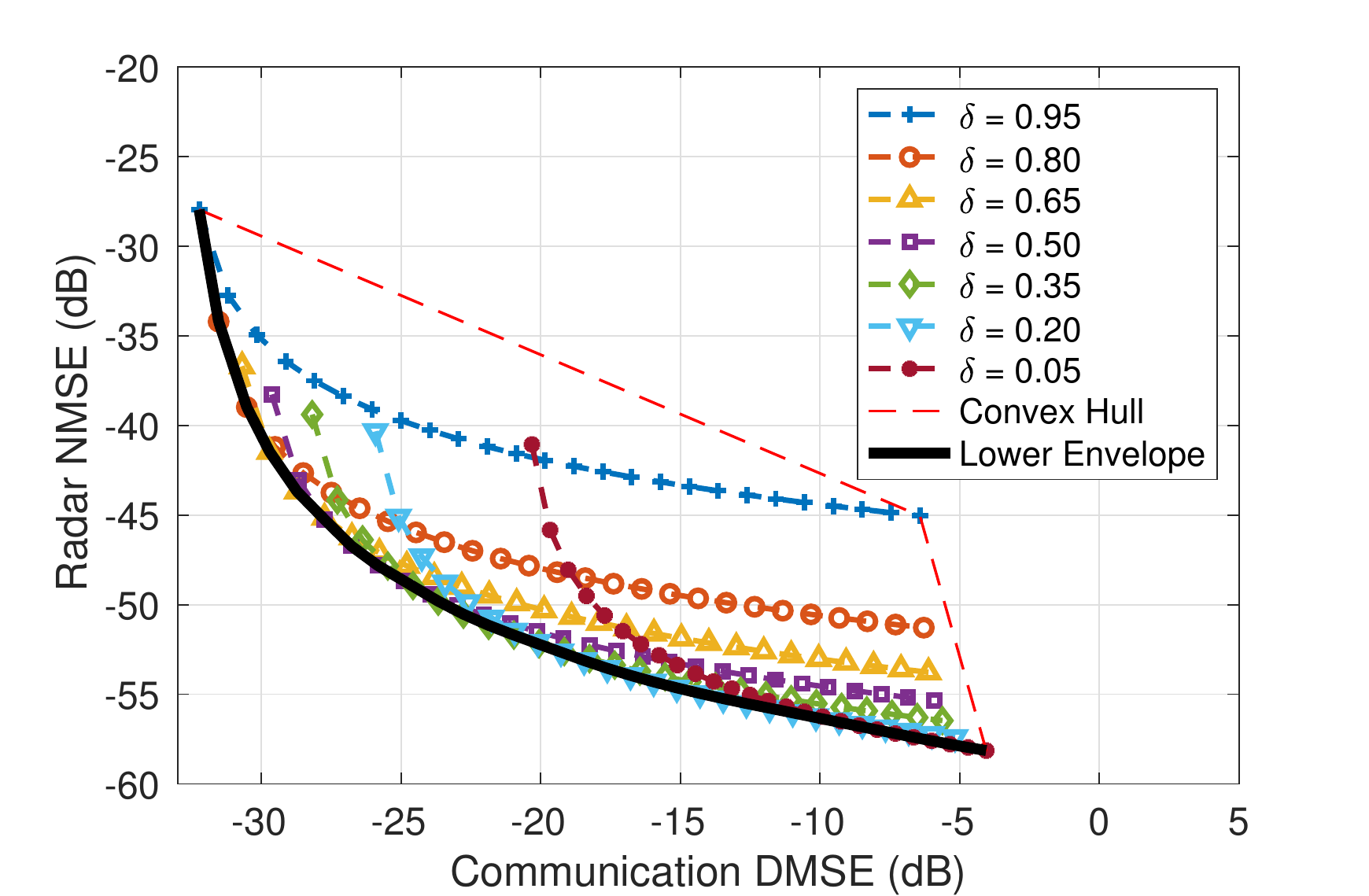}
 \caption{The radar NMSE and the communication DMSE pairs in the JCR
trade-off region with respect to the optimization variable $\rho$ and $\delta$ along with their respective convex hulls at $M=N= 257$, -10 dB SNR, and $K=2$. }
\label{fig:tradeJCR}
\vspace{-1.2em}
\end{figure}

Fig.~\ref{fig:tradeJCR} depict the performance trade-off between the radar NMSE and communication DMSE metrics with respect to the optimization variables $\rho$ and $\delta$ for $M=N= 257$, SNR $\zeta$ of -10 dB, $K=2$, and $\rho = [ 1, 3, \cdots, 53]$. Since $K<<M$, and the SNR is high, the JCR trade-off curve between the radar NMSE versus the communication DMSE is convex for a given $\delta$, as explained in Section~\ref{sec:AdapJCR}. The JCR trade-off curve for a given $\rho$ is almost convex with $\delta$ in logarithmic scale. The deviation from the convex approximation is due to the TX phased-array architecture and the phase shift constraint in the GS algorithm to generate the desired $\bff_\txm(\delta)$. Fig.~\ref{fig:tradeJCR} also illustrates the convex hull of the 2D achievable JCR objective values region, which is the smallest convex set containing the achievable JCR objective values region. The convex hull enables discarding the not so beneficial pairs of the radar NMSE and the communication DMSE in the the 2D feasible JCR achievable objective values region. The lower envelope of the convex hull provides the Pareto-optimal set of the 2D feasible JCR achievable objective values region. 

\begin{figure}[!t]
\begin{minipage}[b]{\linewidth}
  \centering
  \centerline{ \includegraphics[clip,width=\columnwidth]{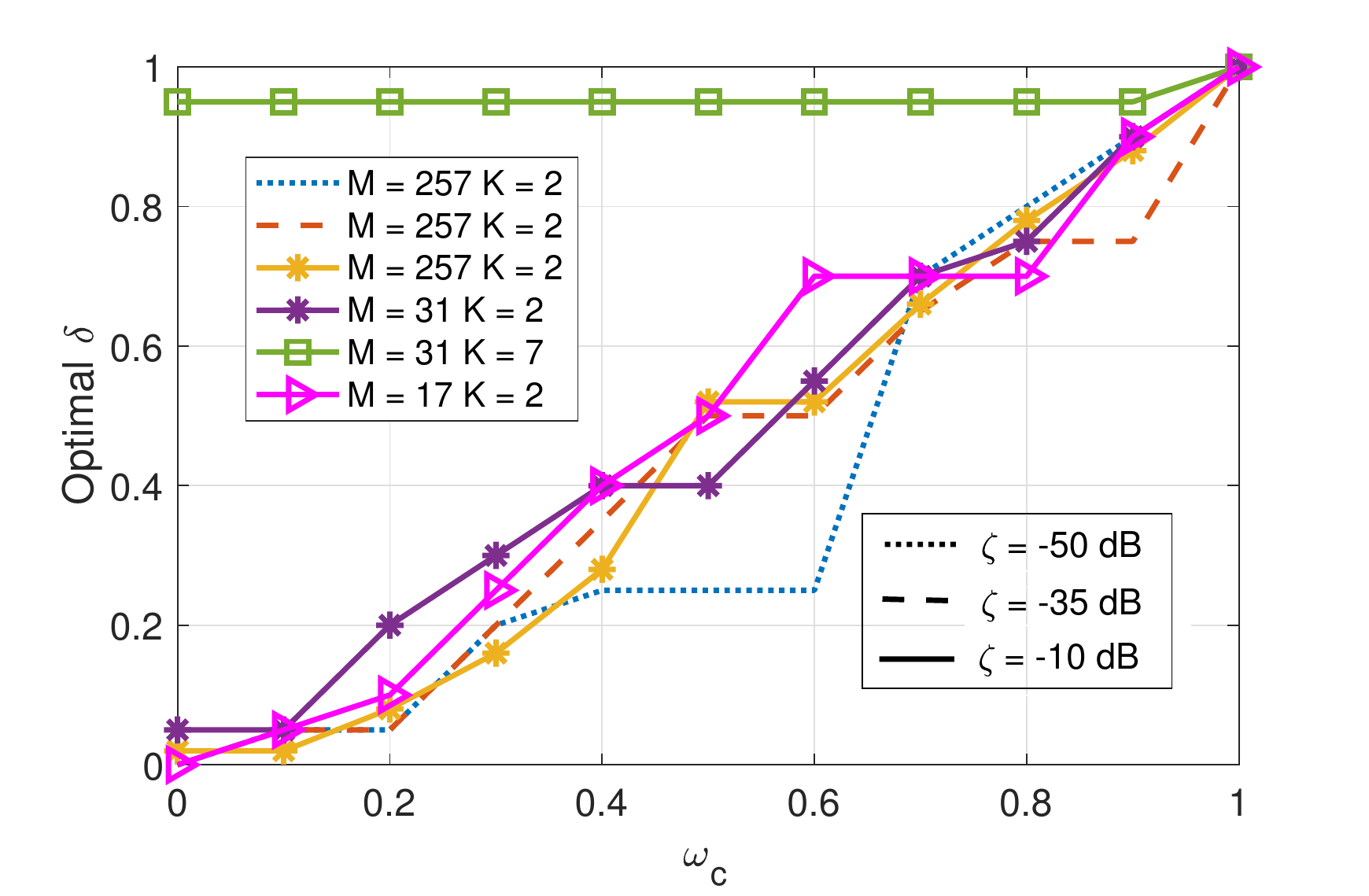}}
  \centerline{(a) Optimal $\delta$ for different weightings}\medskip
\end{minipage}
\begin{minipage}[b]{\linewidth}
  \centering
  \centerline{\includegraphics[clip,width=\columnwidth]{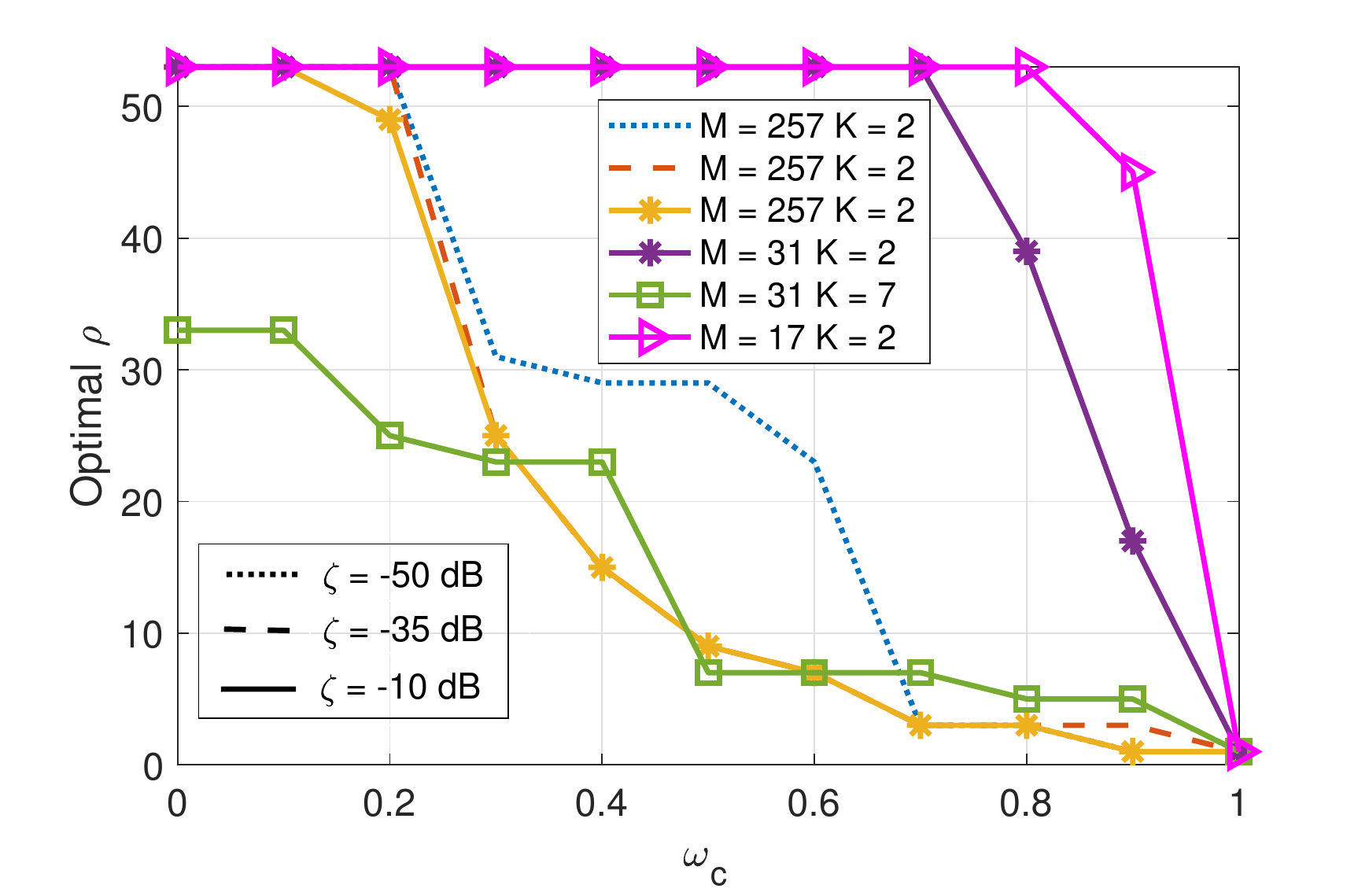}}
  \centerline{(b) Optimal $\rho$ for different weightings}\medskip
\end{minipage}
\caption{Optimal $\delta$ increases and optimal $\rho$ decreases with the communication weightings. The optimal $\optr$ decreases faster than $\delta$ with respect to the $\omega_\mcom$ for large $M$. }
\label{fig:Optw}
\vspace{-1.2em}
\end{figure}

Fig.~\ref{fig:Optw}(a) and (b) show the optimal $\delta$ and $\optr$ versus $\omega_\mcom$ for the optimal weighted average-based JCR design with different $M$ and $K$ at an SNR $\zeta$ of -10 dB, -35 dB, and -50 dB. For small number of targets, $\delta$ increases rapidly with communication weighting. At $\omega_\mcom =1$, $\delta$ converges to 1 with maximum communication spectral efficiency. At $M=31$ and $K=7$, however, the optimal $\delta$ generally remains constant due to the saturation effect and small communication SNR leading to almost linear relation with $\delta$.

\begin{figure}[!ht]
\begin{minipage}[b]{\linewidth}
  \centering
  \centerline{\includegraphics[clip,width=\columnwidth]{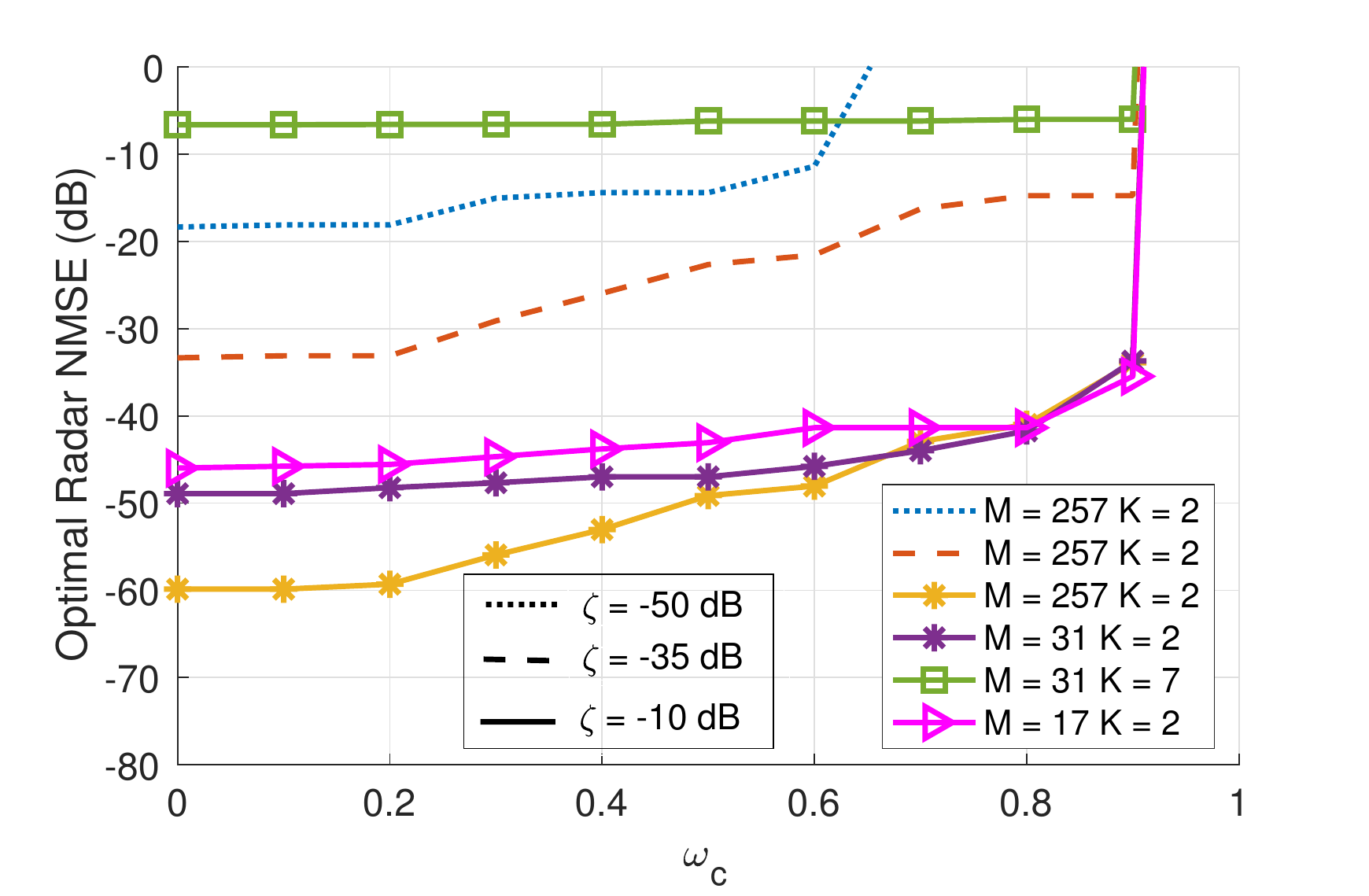}}
  \centerline{(b) Optimal radar NMSE for different weightings}\medskip
\end{minipage}
\begin{minipage}[b]{\linewidth}
  \centering
  \centerline{\includegraphics[clip,width=\columnwidth]{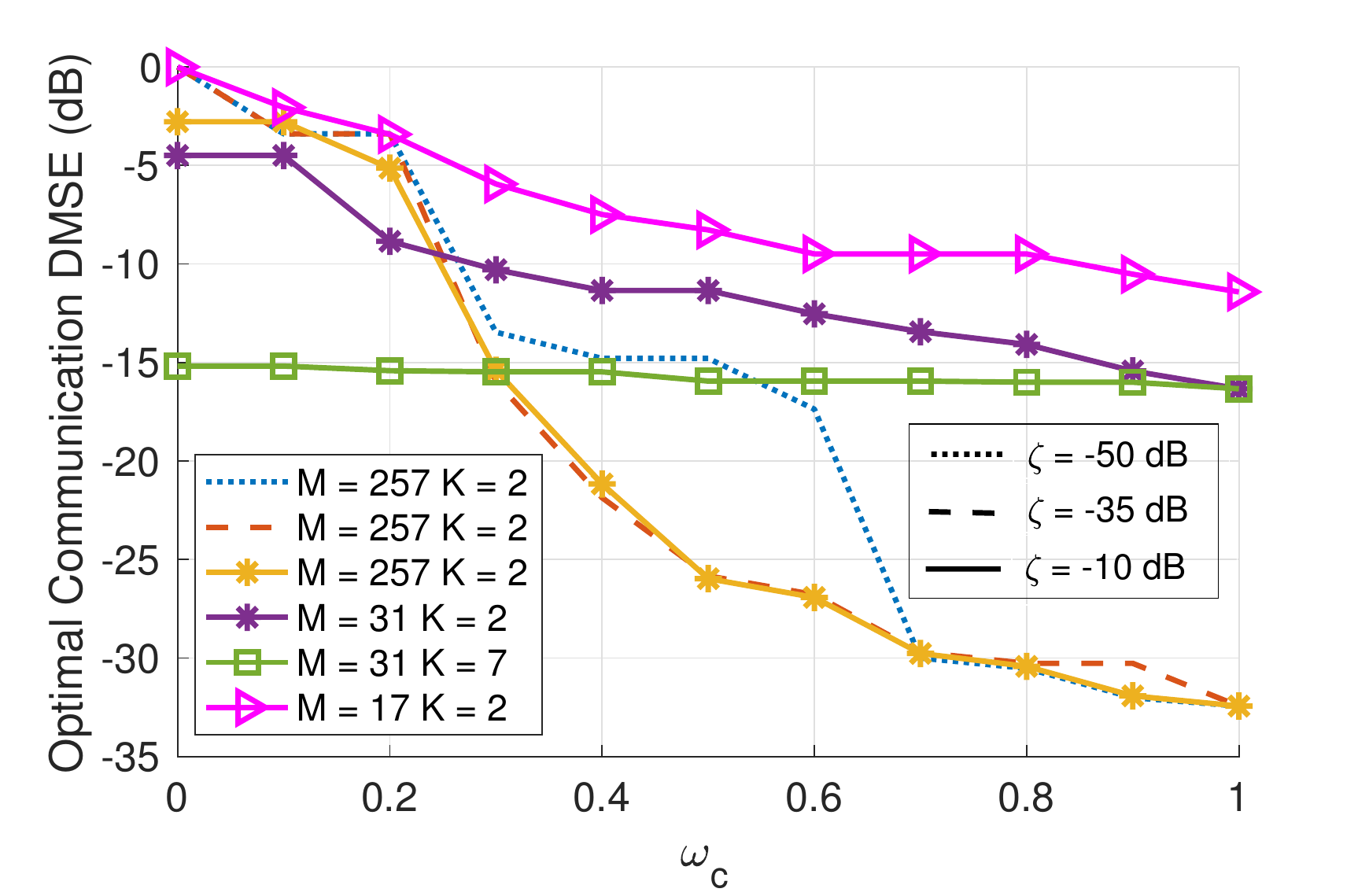}}
  \centerline{(b) Optimal communication DMSE for different weightings}\medskip
\end{minipage}
\caption{Optimal radar NMSE increases and optimal communication DMSE decreases with the communication weightings. This example demonstrates that we can achieve high-resolution SRR/MRR radar channel estimation in the Doppler-angle domain with high accuracy, at the cost of a small reduction in the communication rate.}
\label{fig:OptwRadCom}
\vspace{-1.2em}
\end{figure}

 The optimal $\optr$ decreases faster than $\delta$ with respect to the communication weighting for large $M$. This is due to the fact that communication DMSE degrades much slower with $\delta$, whereas it decreases rapidly with $\optr$. For small $M$, however, optimal $\optr$ is 1 for even large values of communication weighting because $\alpha[\rho]$ doesn't change much with $\optr$. At $\omega_\mcom =1 $, $\rho$ converges to minimum preamble length of 512. At $\omega_\mcom =0$,  $\rho$ converges to $P_\mathrm{max}$ for all the target scenarios, except for $M=31$ and $K=7$ where it converges to a smaller value due to the saturation effect.

Fig.~\ref{fig:OptwRadCom}(a) and (b) show the optimal radar NMSE and communication NMSE versus communication weightings with different $M$ and $K$ at an SNR ($\zeta$) of -10 dB, -35 dB, and -50 dB. The optimal radar NMSE increases with communication weighting, while the optimal communication DMSE decreases with communication weighting. The saturation effect is observed for $M = 31$ and $K = 7$ for both the optimal radar NMSE and the optimal communication DMSE. The radar NMSE goes beyond 0 dB, and is therefore unusable for automotive radar sensing applications, at $\omega_\mcom = 1$ for $\zeta$ of -10 dB and -35 dB, whereas radar is unusable at lower $\omega_\mcom = 0.7$ for  $\zeta$ of -50 dB and $M=257$. The weighted average of the radar NMSE and the communication NMSE decreases with increase in $M$ for $K=2$ and $\zeta = -10$ dB. At $M=257$, the radar NMSE is much better with decreasing $\zeta$, while the communication NMSE is the same all $\zeta$ at low and high communication weightings. This example demonstrates that we can achieve high-resolution radar channel estimation in the Doppler-angle domain with high accuracy, $180^\circ$ field of view, and 5~ms CPI, as desired in the MRR/SRR applications~\cite{PatTorWan:Automotive-Radars:-A-review:17}.

\section{Conclusion and discussion} \label{sec:Conclusion}
In this paper, we proposed an adaptive and fast combined waveform-beamforming design for the mmWave automotive JCR with a phased-array architecture. Our proposed JCR design achieves a wide field of view by transmitting a fraction of energy along the communication direction and distributing the remaining energy ``uniformly" along the other radar sensing directions. Our method uses a few circulant shifts of the designed JCR beamformer and apply 2D partial Fourier CS to rapidly estimate the radar channel in the Doppler-angle domain. To enhance the radar performance, we also optimize these circulant shifts by minimizing the coherence of the compressed sensing matrix under the sampling constraints of the proposed JCR system. Additionally, we develop a MSE-based weighted average optimization-based JCR design with tunable waveform and beamforming parameters that permits a trade-off between the radar NMSE and the communication DMSE metrics.

The results in the paper demonstrate that our proposed JCR design estimated medium- and short-range automotive radar channels in the range-Doppler angle domain with low NMSE and a wide field of view, at the cost of a small reduction in the communication rate. The proposed JCR design with optimized circulant shifts performed better than the random circulant shifts, especially at high SNR and large target counts. Additionally, we observed the saturation effect in both the approaches at high SNR and target counts. The random switching-based JCR design performed very poor as compared to our proposed JCR design due to higher coherence of the resultant CS matrix and the low transmit power under the per-antenna power constraint. In the weighted average optimization-based combined waveform-beamforming JCR design, the optimal preamble length decreases faster than the optimal fraction of communication TX array gain with respect to the communication weightings for large number of frames. The optimal preamble length and the optimal fraction of communication TX array gain remains almost the same for different SNRs at low and high communication weightings.

The results in this paper can be used to develop low-power, small size, spectrum-efficient, and high-performance mmWave devices that will enable next-generation automotive sensing and communication needs. Future work includes an extension of our work for simultaneous range, velocity, angle-of-arrival, and angle-of-departure estimation. It would also be interesting to experimentally evaluate the performance of our proposed JCR design using a measurement platform similar to~\cite{KumMezHea:JCR70:-A-Low-Complexity-Millimeter-Wave:20}.

\section*{Acknowledgment}

The authors would like to thank Prof. Sergiy A. Vorobyov of Aalto University, Finland for discussions regarding the adaptive JCR design.

\bibliographystyle{IEEEtran}
\bibliography{IEEEabrv,refFinal}

\begin{thebibliography}{10}
\providecommand{\url}[1]{#1}
\csname url@samestyle\endcsname
\providecommand{\newblock}{\relax}
\providecommand{\bibinfo}[2]{#2}
\providecommand{\BIBentrySTDinterwordspacing}{\spaceskip=0pt\relax}
\providecommand{\BIBentryALTinterwordstretchfactor}{4}
\providecommand{\BIBentryALTinterwordspacing}{\spaceskip=\fontdimen2\font plus
\BIBentryALTinterwordstretchfactor\fontdimen3\font minus
  \fontdimen4\font\relax}
\providecommand{\BIBforeignlanguage}[2]{{%
\expandafter\ifx\csname l@#1\endcsname\relax
\typeout{** WARNING: IEEEtran.bst: No hyphenation pattern has been}%
\typeout{** loaded for the language `#1'. Using the pattern for}%
\typeout{** the default language instead.}%
\else
\language=\csname l@#1\endcsname
\fi
#2}}
\providecommand{\BIBdecl}{\relax}
\BIBdecl

\bibitem{PatTorWan:Automotive-Radars:-A-review:17}
S.~M. Patole, M.~Torlak, D.~Wang, and M.~Ali, ``Automotive radars: A review of
  signal processing techniques,'' \emph{{IEEE} Signal Process. Mag.}, vol.~34,
  no.~2, pp. 22--35, March 2017.

\bibitem{ChoVaGon:Millimeter-Wave-Vehicular-Communication:16}
J.~Choi, V.~Va, N.~Gonzalez-Prelcic, R.~Daniels, C.~R. Bhat, and R.~W. Heath,
  ``Millimeter-wave vehicular communication to support massive automotive
  sensing,'' \emph{{IEEE} Commun. Mag.}, vol.~54, no.~12, pp. 160--167, Dec.
  2016.

\bibitem{ieee2012wireless}
``{Wireless LAN Medium Access Control (MAC) and Physical Layer (PHY)
  Specifications. Amendment 3: Enhancements for Very High Throughput in the 60
  GHz Band},'' \emph{IEEE Std. 802.11ad}, 2012.

\bibitem{MisBhaKoi:Toward-Millimeter-Wave-Joint:19}
K.~V. {Mishra}, M.~R. {Bhavani Shankar}, V.~{Koivunen}, B.~{Ottersten}, and
  S.~A. {Vorobyov}, ``Toward millimeter-wave joint radar communications: {A}
  signal processing perspective,'' \emph{IEEE Signal Process. Mag.}, vol.~36,
  no.~5, pp. 100--114, Sep. 2019.

\bibitem{JRC2020}
D.~{Ma}, N.~{Shlezinger}, T.~{Huang}, Y.~{Liu}, and Y.~C. {Eldar}, ``Joint
  radar-communication strategies for autonomous vehicles: Combining two key
  automotive technologies,'' \emph{IEEE Signal Process. Mag.}, vol.~37, no.~4,
  pp. 85--97, Jul. 2020.

\bibitem{DokShaSti:Multicarrier-Phase-Modulated:18}
S.~H. Dokhanchi, M.~R.~B. Shankar, T.~Stifter, and B.~Ottersten, ``Multicarrier
  phase modulated continuous waveform for automotive joint radar-communication
  system,'' in \emph{Proc. Int. Workshop on Signal Process. Advances in
  Wireless Commun.}, Jun. 2018, pp. 1--5.

\bibitem{HasAmiZha:Signaling-strategies-for-dual-function:16}
A.~{Hassanien}, M.~G. {Amin}, Y.~D. {Zhang}, and F.~{Ahmad}, ``Signaling
  strategies for dual-function radar communications: an overview,'' \emph{IEEE
  Aerosp. Electron. Sys. Mag.}, vol.~31, no.~10, pp. 36--45, Oct 2016.

\bibitem{KumChoGon:IEEE-802.11ad-Based-Radar::18}
P.~Kumari, J.~Choi, N.~Gonz{\'a}lez-Prelcic, and R.~W. Heath, ``{IEEE}
  802.11ad-based radar: An approach to joint vehicular communication-radar
  system,'' \emph{{IEEE} Trans. Veh. Technol.}, vol.~67, no.~4, pp. 3012--3027,
  Apr. 2018.

\bibitem{GroLopVen:Opportunistic-Radar-in-IEEE:18}
E.~Grossi, M.~Lops, L.~Venturino, and A.~Zappone, ``Opportunistic radar in
  {IEEE 802.11ad} networks,'' \emph{{IEEE} Trans. Signal Process.}, vol.~66,
  no.~9, pp. 2441--2454, May 2018.

\bibitem{MunMisGue:Beam-Alignment-and-Tracking:19}
G.~R. {Muns}, K.~V. {Mishra}, C.~B. {Guerra}, Y.~C. {Eldar}, and K.~R.
  {Chowdhury}, ``Beam alignment and tracking for autonomous vehicular
  communication using ieee 802.11ad-based radar,'' in \emph{Proc. IEEE Conf. on
  Comput. Commun. Workshops (INFOCOM WKSHPS)}, Apr. 2019, pp. 535--540.

\bibitem{KumEltHea:Sparsity-aware-adaptive-beamforming:18}
P.~{Kumari}, M.~E. {Eltayeb}, and R.~W. {Heath}, ``Sparsity-aware adaptive
  beamforming design for {IEEE 802.11ad}-based joint communication-radar,'' in
  \emph{Proc. IEEE Radar Conf.}, April 2018, pp. 0923--0928.

\bibitem{KumMazMez:Low-Resolution-Sampling-for-Joint:18}
P.~Kumari, K.~U. Mazher, A.~Mezghani, and R.~W. Heath, ``Low resolution
  sampling for joint millimeter-wave {MIMO} communication-radar,'' in
  \emph{Proc. IEEE Statistical Signal Process. Workshop}, Jun. 2018, pp.
  193--197.

\bibitem{LiJosTao:Feasibility-study-on-full-duplex:14}
L.~Li, K.~Josiam, and R.~Taori, ``Feasibility study on full-duplex wireless
  millimeter-wave systems,'' May 2014, pp. 2769--2773.

\bibitem{KumVorHea:Adaptive-Virtual-Waveform:20}
P.~{Kumari}, S.~A. {Vorobyov}, and R.~W. {Heath}, ``Adaptive virtual waveform
  design for millimeter-wave joint communication--radar,'' \emph{IEEE Trans. on
  Signal Process.}, vol.~68, pp. 715--730, Nov. 2020.

\bibitem{Rau:Compressive-sensing-and-structured:10}
H.~Rauhut, ``Compressive sensing and structured random matrices,''
  \emph{Theoretical foundations and numerical methods for sparse recovery},
  vol.~9, pp. 1--92, 2010.

\bibitem{KumMyeVor:A-Combined-Waveform-Beamforming-Design:19}
P.~{Kumari}, N.~J. {Myers}, S.~A. {Vorobyov}, and R.~W. {Heath}, ``A combined
  waveform-beamforming design for millimeter-wave joint communication-radar,''
  in \emph{Proc. Asilomar Conf. Signals, Syst., and Comput.}, 2019, pp.
  1422--1426.

\bibitem{MyeMezHea:FALP:-Fast-beam:19}
N.~J. {Myers}, A.~{Mezghani}, and R.~W. {Heath}, ``{FALP}: Fast beam alignment
  in {mmWave} systems with low-resolution phase shifters,'' \emph{IEEE Trans.
  Commun.}, pp. 1--1, 2019.

\bibitem{Ger:A-practical-algorithm-for-the-determination:72}
R.~W. Gerchberg, ``A practical algorithm for the determination of phase from
  image and diffraction plane pictures,'' \emph{Optik}, vol.~35, pp. 237--246,
  1972.

\bibitem{HeaGonRan:An-Overview-of-Signal-Processing:16}
R.~W. {Heath}, N.~{Gonz{\'a}lez-Prelcic}, S.~{Rangan}, W.~{Roh}, and A.~M.
  {Sayeed}, ``An overview of signal processing techniques for millimeter wave
  mimo systems,'' \emph{IEEE J. Sel. Topics Signal Process.}, vol.~10, no.~3,
  pp. 436--453, 2016.

\bibitem{JunZhoCha:Beam-codebook-based:09}
{Junyi Wang}, {Zhou Lan}, {Chang-woo Pyo}, T.~{Baykas}, {Chin-sean Sum}, M.~A.
  {Rahman}, {Jing Gao}, R.~{Funada}, F.~{Kojima}, H.~{Harada}, and S.~{Kato},
  ``Beam codebook based beamforming protocol for {multi-Gbps} millimeter-wave
  {WPAN} systems,'' \emph{IEEE J. Sel. Areas Commun.}, vol.~27, no.~8, pp.
  1390--1399, 2009.

\bibitem{li2012convolutional}
K.~Li, L.~Gan, and C.~Ling, ``Convolutional compressed sensing using
  deterministic sequences,'' \emph{IEEE Trans. on Signal Process.}, vol.~61,
  no.~3, pp. 740--752, 2012.

\bibitem{LusDonSan:Compressed-Sensing-MRI:08}
M.~{Lustig}, D.~L. {Donoho}, J.~M. {Santos}, and J.~M. {Pauly}, ``Compressed
  sensing {MRI},'' \emph{IEEE Signal Process. Mag.}, vol.~25, no.~2, pp.
  72--82, 2008.

\bibitem{CanEldNee:Compressed-sensing-with:11}
E.~J. Candes, Y.~C. Eldar, D.~Needell, and P.~Randall, ``Compressed sensing
  with coherent and redundant dictionaries,'' \emph{Appl. Computat. Harmon.
  Anal.}, vol.~31, no.~1, pp. 59--73, 2011.

\bibitem{LusDonPau:Sparse-MRI:-The-application:07}
M.~Lustig, D.~Donoho, and J.~M. Pauly, ``Sparse {MRI}: The application of
  compressed sensing for rapid {MR} imaging,'' \emph{Magnetic Resonance in
  Medicine: An Official Journal of the International Society for Magnetic
  Resonance in Medicine}, vol.~58, no.~6, pp. 1182--1195, 2007.

\bibitem{PatEasHea:Compressed-synthetic-aperture:10}
V.~M. Patel, G.~R. Easley, D.~M. Healy~Jr, and R.~Chellappa, ``Compressed
  synthetic aperture radar,'' \emph{IEEE J. Sel. Topics Signal Process.},
  vol.~4, no.~2, pp. 244--254, 2010.

\bibitem{Chu:Polyphase-codes-with:72}
D.~Chu, ``Polyphase codes with good periodic correlation properties
  (corresp.),'' \emph{IEEE Trans. Inf. Theory}, vol.~18, no.~4, pp. 531--532,
  1972.

\bibitem{Luk:Sequences-and-arrays-with:88}
H.~D. Luke, ``Sequences and arrays with perfect periodic correlation,''
  \emph{IEEE Trans. Aerosp. Electron. Syst.}, vol.~24, no.~3, pp. 287--294,
  1988.

\bibitem{CaiWan:Orthogonal-matching-pursuit:11}
T.~T. Cai and L.~Wang, ``Orthogonal matching pursuit for sparse signal recovery
  with noise,'' \emph{IEEE Trans. Inf. Theory}, vol.~57, no.~7, pp. 4680--4688,
  Jun. 2011.

\bibitem{TanWuHer:Performance-Analysis-of-OMP-Based:18}
G.~{Tan}, B.~{Wu}, and T.~{Herfet}, ``Performance analysis of {OMP}-based
  channel estimations in mobile {OFDM} systems,'' \emph{IEEE Trans. on Wireless
  Commun.}, vol.~17, no.~5, pp. 3459--3473, 2018.

\bibitem{HeaLoz:Foundations-of-MIMO-Communication:19}
R.~W. Heath~Jr and A.~Lozano, \emph{Foundations of MIMO Communication}.\hskip
  1em plus 0.5em minus 0.4em\relax Cambridge University Press, 2019.

\bibitem{CovTho:Elements-of-information-theory:12}
T.~M. Cover and J.~A. Thomas, \emph{Elements of information theory}.\hskip 1em
  plus 0.5em minus 0.4em\relax John Wiley \& Sons, 2012.

\bibitem{Bli:Cooperative-radar-and-communications:14}
D.~W. Bliss, ``Cooperative radar and communications signaling: The estimation
  and information theory odd couple,'' in \emph{Proc. IEEE Radar Conf.}, May
  2014, pp. 50--55.

\bibitem{brehmer2012utility}
J.~Brehmer, \emph{Utility Maximization in Nonconvex Wireless Systems}.\hskip
  1em plus 0.5em minus 0.4em\relax Springer Science \& Business Media, 2012,
  vol.~5.

\bibitem{BoyVan:Convex-optimization:04}
S.~Boyd and L.~Vandenberghe, \emph{Convex optimization}.\hskip 1em plus 0.5em
  minus 0.4em\relax Cambridge University Press, 2004.

\bibitem{HasTopSch:Millimeter-wave-technology-for-automotive:12}
J.~Hasch, E.~Topak, R.~Schnabel, T.~Zwick, R.~Weigel, and C.~Waldschmidt,
  ``{Millimeter-wave technology for automotive radar sensors in the 77 {GHz}
  frequency band},'' \emph{IEEE Trans. Microw. Theory Techn.}, vol.~60, no.~3,
  pp. 845--860, 2012.

\bibitem{KumMezHea:JCR70:-A-Low-Complexity-Millimeter-Wave:20}
P.~Kumari, A.~Mezghani, and R.~W. Heath~Jr, ``{JCR70}: A low-complexity
  millimeter-wave proof-of-concept platform for a fully-digital {MIMO} joint
  communication-radar,'' \emph{arXiv preprint arXiv:2006.13344}, Jun. 2020.

\end{thebibliography}

\end{document}